\documentclass[12pt]{article}
\usepackage{amsmath}
\usepackage{amsthm}
\usepackage{amssymb}
\usepackage{dsfont}
\usepackage{algorithm}
\usepackage{booktabs}
\usepackage{multirow}
\usepackage{fullpage}
\usepackage{setspace}
\usepackage{hyperref}

\newcommand{\vertiii}[1]{{\|\kern-0.2ex| #1
\|\kern-0.2ex|}}
\usepackage{graphicx}
\newcommand{\argmin}{\operatorname*{arg \ min}}
\newcommand{\minim}{\operatorname*{minimize}}
\newcommand{\mbOmega}{\boldsymbol{\Omega}}
\newcommand{\bovtex}{\boldsymbol}
\newcommand{\mbomega}{\boldsymbol{\omega}}

\newcommand{\mbTheta}{\boldsymbol{\Theta}}

\def\@pnumwidth{21pt}
\usepackage{natbib}
\newtheorem{prop}{Proposition}

\theoremstyle{plain}
\newtheorem{thm}{Theorem}[section]

\begin{document}

\title{Direct covariance matrix estimation with\\ compositional data}
\author{
Aaron J. Molstad$^{*}$, Karl Oskar Ekvall$^{\star,\dagger}$, and Piotr M. Suder$^{\ddagger}$\medskip\\
School of Statistics, University of Minnesota$^*$\\
Department of Statistics, University of Florida$^\star$\\
Division of Biostatistics, Institute of Environmental Medicine,\\Karolinska Institutet$^\dagger$\\
Department of Statistical Science, Duke University$^\ddagger$
}

\maketitle
\begin{abstract}
Compositional data arise in many areas of research in the natural and biomedical
sciences. One prominent example is in the study of the human gut microbiome,
where researchers can measure the relative abundance of many distinct microorganisms. Often, practitioners are interested in learning how
the dependencies between microbes vary across distinct populations or experimental
conditions. In statistical terms, the goal is to estimate a covariance
matrix for the (latent) log-abundances of the microbes in each of the populations.
However, the compositional nature of the data prevents the use of standard
estimators for these covariance matrices. In this article, we propose an
estimator of multiple covariance matrices which allows for information
sharing across distinct populations of samples. Compared to some existing
estimators, which estimate the covariance matrices of interest indirectly,
our estimator is direct, ensures positive definiteness, and is the solution
to a convex optimization problem. We compute our estimator using a proximal-proximal
gradient descent algorithm. Asymptotic properties of our estimator reveal
that it can perform well in high-dimensional settings. We show that our
method provides more reliable estimates than competitors in an analysis
of microbiome data from subjects with myalgic encephalomyelitis/chronic
fatigue syndrome and through simulation studies. 
\end{abstract}




\section{Introduction}
\label{sec:Introduction}

High-dimensional compositional data arise in many areas of modern science.
To study the human gut microbiome, for example, practitioners measure the
relative abundances of various microbes using next-generation sequencing
followed by alignment and normalization \citep{gloor2017microbiome}. For
each subject in a study, the resulting measurement is a $p$-dimensional
vector which has nonnegative entries and sums to one
\citep{huson2007megan,
segata2012metagenomic}. More generally, compositional data arise when,
for example, one observes multivariate count-valued data wherein the total counts in
a sample is an experimental artifact. Here, we focus on compositional data
which belong to the set
\begin{equation*}
\mathbb{C}^{p-1} = \left \{x \in \mathbb{R}^{p}: \sum _{j=1}^{p} x_{j}
= 1, x_{j} > 0 \text{ for each }j \in [p]\right \},
\end{equation*}
where $[p] = \{1, \dots , p\}$ for a positive integer $p$.

To make matters concrete, let
$X = (X_{1}, \dots , X_{p})^{\top }\in \mathbb{C}^{p-1}$ be a random composition
whose components correspond to the variables of interest. Letting
$W = (W_{1}, \dots , W_{p})^{\top}$ denote the corresponding latent abundances,
also known as the basis \citep{aitchison1982statistical}, we assume
\begin{equation*}
X_{j} = \frac{W_{j}}{\sum _{k=1}^{p} W_{k}}, ~~~ j \in [p],
\end{equation*}
where each $W_{j} \in (0,\infty )$. When characterizing the dependence
between any two components from the compositional vector, the parameter
of interest is often the basis covariance matrix
$\Omega ^{*} \in \mathbb{S}^{p}_{+}$, where
\begin{equation*}
\Omega ^{*}_{jk} = {\mathrm{Cov}}\left \{\log (W_{j}), \log (W_{k})\right
\},~~~(j,k) \in [p] \times [p],
\end{equation*}
and $\mathbb{S}^{p}_{+}$ denotes the set of $p \times p$ symmetric positive
definite matrices.

In many studies involving compositional microbiome data, practitioners
are interested in modeling the interactions and dependencies between microbe
abundance \citep{faust2012microbial,ma2021networks}. For instance, one
may want to estimate whether two microbes occur in higher frequencies jointly.
The basis covariance matrix $\Omega ^{*}$ provides one route for addressing
such questions
\citep{jiang2019microbiome,matchado2021network,he2021robust}, but is not
straightforward to estimate from independent realizations of $X$ because
$W$ is latent. One common approach relies on the estimation of the variation
matrix $\Theta ^{*}$ \citep[Chapter
4]{aitchison2003statistical}, defined elementwise by
\begin{align*}
\Theta ^{*}_{jk} & = {\mathrm{Var}}\left \{ \log (X_{j}/X_{k})\right \},
\\
& = {\mathrm{Var}}\left \{ \log (W_{j}) - \log (W_{k})\right \}
\\
& = {\mathrm{Var}}\left \{\log (W_{j})\right \} + {\mathrm{Var}}\left \{\log (W_{k})
\right \} - 2 {\mathrm{Cov}}\left \{ \log (W_{j}), \log (W_{k})\right \}.
\end{align*}
Thus, letting
$\omega ^{*} = {\mathrm{diag}}(\Omega ^{*}) \in \mathbb{R}^{p}$ and
$\mathds{1}_{p} = (1,1,\dots ,1)^{\top }\in \mathbb{R}^{p}$,
%
\begin{equation}
\label{eq:variationMatrix}
\Theta ^{*} = \omega ^{*} \mathds{1}_{p}^{\top }+ \mathds{1}_{p}
\omega ^{*\top} - 2 \Omega ^{*}.
\end{equation}
To define an estimator of $\Theta ^{*}$, let
$x_{i} = (x_{i1}, \dots , x_{ip})^{\top }\in \mathbb{C}^{p-1}$,
$i\in [n]$, denote independent realizations of $X$. Let also
$z_{ijk} = \log (x_{ij}/x_{ik})$ and
$\bar{z}_{jk} = n^{-1} \sum _{i=1}^{n} z_{ijk}$ for all
$(j,k) \in [p]\times [p]$. The sample estimator $\widehat{\Theta}$ is defined
elementwise by
\begin{equation*}
\widehat{\Theta}_{jk} = \frac{1}{n}\sum _{i=1}^{n} (z_{ijk} - \bar{z}_{jk})^{2}.
\end{equation*}

While $\widehat{\Theta}$ is a natural estimator of $\Theta ^{*}$, it is
unclear how to use it to estimate $\Omega ^{*}$ in general because there
are infinitely many $\widehat\Omega $ such that
$\widehat{\Theta} = \widehat\omega \mathds{1}_{p}^{\top }+ \mathds{1}_{p}
\widehat\omega ^{\top} - 2 \widehat\Omega $. Namely, the diagonal entries
of $\widehat{\Theta}$ and
$\widehat\omega \mathds{1}_{p}^{\top }+ \mathds{1}_{p}
\widehat\omega ^{\top} - 2 \widehat\Omega $ are zero, so there are
$p(p-1)/2$ unique equalities but $p(p+1)/2$ unknowns in
$\widehat{\Omega}$. However, if one assumes that many entries of
$\Omega ^{*}$ are zero, then it can be estimated based on~\eqref{eq:variationMatrix}. \citet{cao2016large} proved that if
$p \geq 5$ and $\Omega ^{*}$ has fewer than $(p-1)$ nonzero off-diagonal
entries, then no two $\Omega ^{*}$ correspond to the same
$\Theta ^{*}$. Thus, if one could assume $s < p-1$ off-diagonal entries
of $\Omega ^{*}$ are nonzero, one could consider the estimator
%
\begin{equation}
\label{eq:L0_estimator}
\argmin _{\Omega = \Omega ^{\top}} \|\widehat{\Theta} - \omega
\mathds{1}_{p}^{\top }- \mathds{1}_{p} \omega ^{\top }+ 2 \Omega \|_{F}^{2}~~
\text{subject to }\|\Omega ^{-}\|_{0} \leq s,
\end{equation}
where $\Omega ^{-}$ denotes the matrix $\Omega $ with its diagonal entries
set to zero,
$\|A\|_{F}^{2} = {\mathrm{tr}}(A^{\top }A) = \sum _{j,k}A_{jk}^{2} $ is the
squared Frobenius norm of a matrix $A$, and
$\|A\|_{0} = \sum _{j,k} \mathbf{1}(A_{jk} \neq 0)$ counts
the number of nonzero entries in $A$. In practice, assuming only
$s < p - 1$ off-diagonal elements are non-zero is often too restrictive.
Of course,~\eqref{eq:L0_estimator} could also be used with
$s \geq p-1$, but due to the $L_{0}$ constraint,~\eqref{eq:L0_estimator} is the solution to a nonconvex optimization problem
and is computationally challenging for large $p$.

In view of~\eqref{eq:L0_estimator} and its limitations, and given that
the $L_{1}$ norm is a convex relaxation of the $L_{0}$ norm, a natural
alternative is
%
\begin{equation}
\label{eq:L1_estimator}
\argmin _{\Omega = \Omega ^{\top}} \left \{ \|\widehat{\Theta} -
\omega \mathds{1}_{p}^{\top }- \mathds{1}_{p} \omega ^{\top }+ 2
\Omega \|_{F}^{2} + \lambda \|\Omega ^{-}\|_{1}\right \},
\end{equation}
where $\|A\|_{1} = \sum _{j,k} |A_{jk}|$ for a matrix $A$.
\citet{cao2016large} described their estimator as a ``one-step approximation
to~\eqref{eq:L1_estimator}'', but did not study~\eqref{eq:L1_estimator}. Appealingly, the problem in~\eqref{eq:L1_estimator} can be recast as an $L_{1}$-penalized least squares
problem and computed via existing algorithms. However, neither~\eqref{eq:L0_estimator},~\eqref{eq:L1_estimator}, nor the method of
\citet{cao2016large} provide estimates which are guaranteed to be positive
definite, or even nonnegative definite (see Section~\ref{subsec:PD}). Replacing
the feasible set in~\eqref{eq:L0_estimator} or~\eqref{eq:L1_estimator} by $\mathbb{S}^{p}_{+}$, or a subset thereof, complicates
computation substantially. For example, even in the context of standard
covariance matrix estimation (i.e., when the $\log (W_{j})$ are observable),
enforcing sparsity and positive definiteness simultaneously is challenging
\citep{bien2011sparse,rothman2012positive,xue2012positive,XuProximal}.

In many applications, one requires a basis covariance matrix estimate from
multiple distinct populations. For example, in our motivating data analysis,
the goal is to compare how the microbes interact in the gut of patients
with myalgic encephalomyelitis/chronic fatigue syndrome (ME/CFS) versus
controls \citep{Giloteaux2016}. To estimate the two basis covariance matrices,
one could apply existing estimators to each of the populations (ME/CFS
patients and controls) separately. However, sample sizes are often small
relative to the dimension of the basis covariance. For example, there are
only 37 and 47 control and ME/CFS patients, respectively, used to estimate
both $39 \times 39$ basis covariance matrices.

\begin{figure}[t!]
\begin{center}
\includegraphics[width=6.0cm]{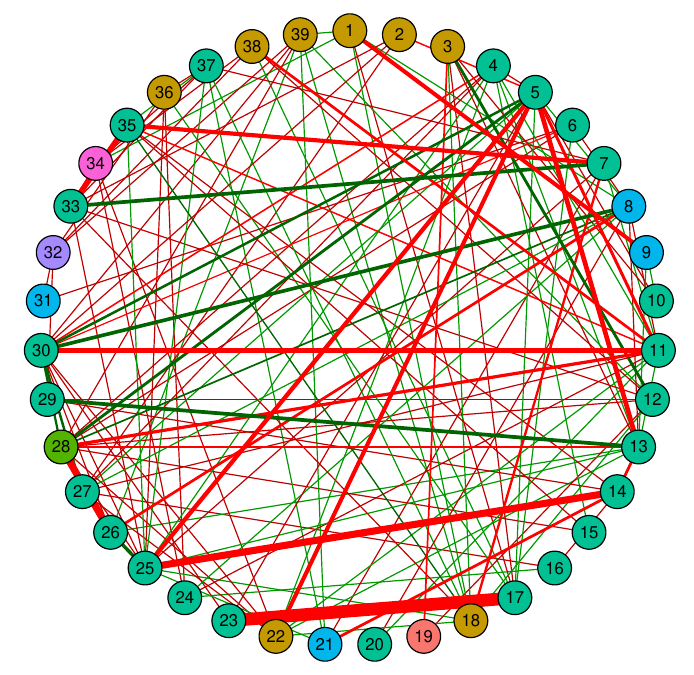}
~~~~~~~~~~\includegraphics[width=7.65cm]{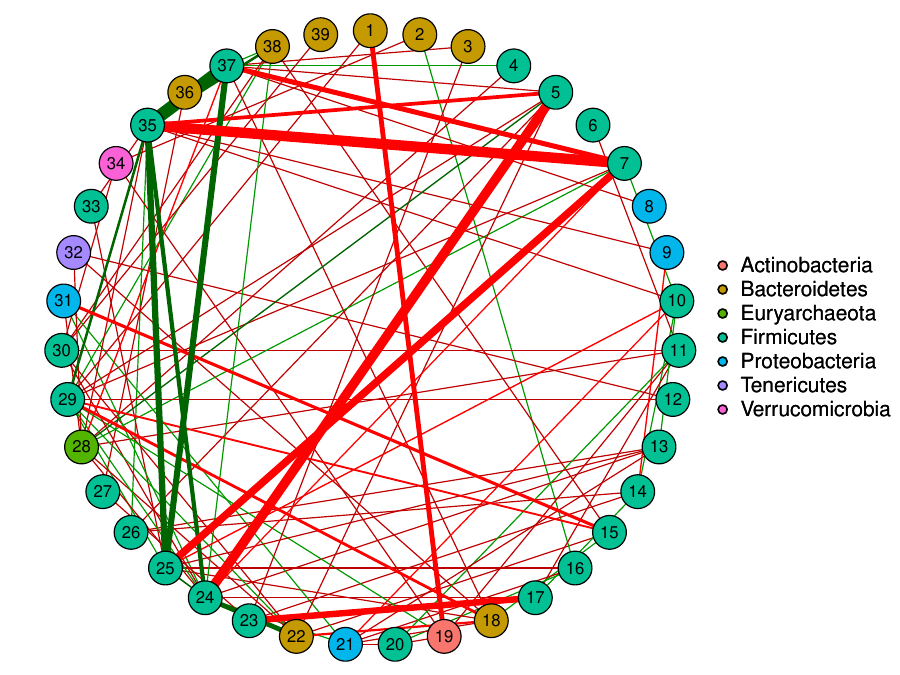}\\
(a) Controls~~~~~~~~~~~~~~~~~~~~~~~~~~~~~~~~~~~~~~~~~(b) ME/CFS~~~~~~~~~~~~
\end{center}
\caption{Estimated correlation networks for controls and patients with ME/CFS
\citep{Giloteaux2016} using the method of \citet{cao2016large}. Each node
corresponds to an OTU as described in Section
\protect \ref{sec:real_data_anaylsis}. Green edges denote positive estimated correlations,
red edges denote negative estimated correlations, and the absence of an
edge indicates an estimated correlation of zero. The thickness of an edge
indicates the magnitude of the correlation: thicker corresponds to a larger
magnitude. Panel (a) is the network estimated from control patients while
panel (b) is the network estimated from patients with ME/CFS.}%
\label{fig:estimated_covariances_chronicfatigue_coat}
\end{figure}

A more efficient approach would estimate the two covariance matrices jointly
in order to borrow information across populations. If, for instance, the
basis covariances have similar sparsity patterns, exploiting this shared
information across populations can substantially improve efficiency. Joint
estimation is especially common in the literature on estimating sparse
covariance and inverse covariance matrices from multivariate normal data
collected on multiple populations
\citep{bigot2011group, guo2011joint,danaher2014joint,price2015ridge,ma2016joint,cai2016joint,saegusa2016joint, price2021estimating}.
In the context of estimating basis covariance matrices from microbiome
data, it is natural to assume the covariance matrices have similar sparsity
patterns. Biologically, it is often reasonable to assume there are microbes
whose abundances are uncorrelated in all the populations in a study. For
example, in Section~\ref{sec:real_data_anaylsis}, when we estimate the
basis covariance matrices for ME/CFS patients and controls using our method,
which shares information across populations, we estimate identical sparsity
patterns. In contrast, when we estimate these matrices separately using
an existing method, few estimated nonzero correlations are shared between
populations (Figure~\ref{fig:estimated_covariances_chronicfatigue_coat}). We investigate the
reliability of these estimates in Section~\ref{subsec:stability}.

In this article, we study~\eqref{eq:L1_estimator} under positive definite
constraints, and propose a generalization of~\eqref{eq:L1_estimator} for
estimating multiple covariance matrices simultaneously. We establish asymptotic
error bounds for both the single and multiple population versions of our
estimator, and we propose an efficient algorithm for their computation.
In simulation studies and our analysis of the ME/CFS microbiome data, we
demonstrate that our methods can provide more reliable estimates of the
covariance matrices of interest than existing competitors.

\section{Methodology}
\label{sec2}

\subsection{Multiple basis covariance matrix estimation}
\label{sec2.1}

In the remainder of this article, we let the subscript $(h)$ denote data
or population parameters from the $h$th population, $h \in [H]$ for some
$H \geq 1$. For example, $x_{(h)i} \in \mathbb{R}^{p}$ is the vector with
compositional data for observation $i \in [n_{(h)}]$ in the $h$th population.
Similarly, $\Omega ^{*}_{(h)}$ is the basis covariance for the $h$th population.

We focus on estimating
$\Omega ^{*}_{(1)}, \dots , \Omega ^{*}_{(H)}$ using the data
$\{x_{(h)i} \in \mathbb{R}^{p}: h \in [H], i\in n_{(h)}\}$. As argued in
Section~\ref{sec:Introduction}, one can estimate any
$\Theta ^{*}_{(h)}$ using
\begin{equation*}
\widehat{\Theta}_{(h)jk} = \frac{1}{n_{(h)}} \sum _{i=1}^{n_{(h)}}(z_{(h)ijk}
- \bar{z}_{(h)jk})^{2}, ~~~(j,k) \in [p]\times [p],
\end{equation*}
where $z_{(h)ijk} = \log (x_{(h)ij}/x_{(h)ik})$ and
$\bar{z}_{(h)jk} = n_{(h)}^{-1} \sum _{i=1}^{n_{(h)}} z_{(h)ijk}$.

\begin{figure}[t!]
\begin{center}
\includegraphics[width=0.9\textwidth]{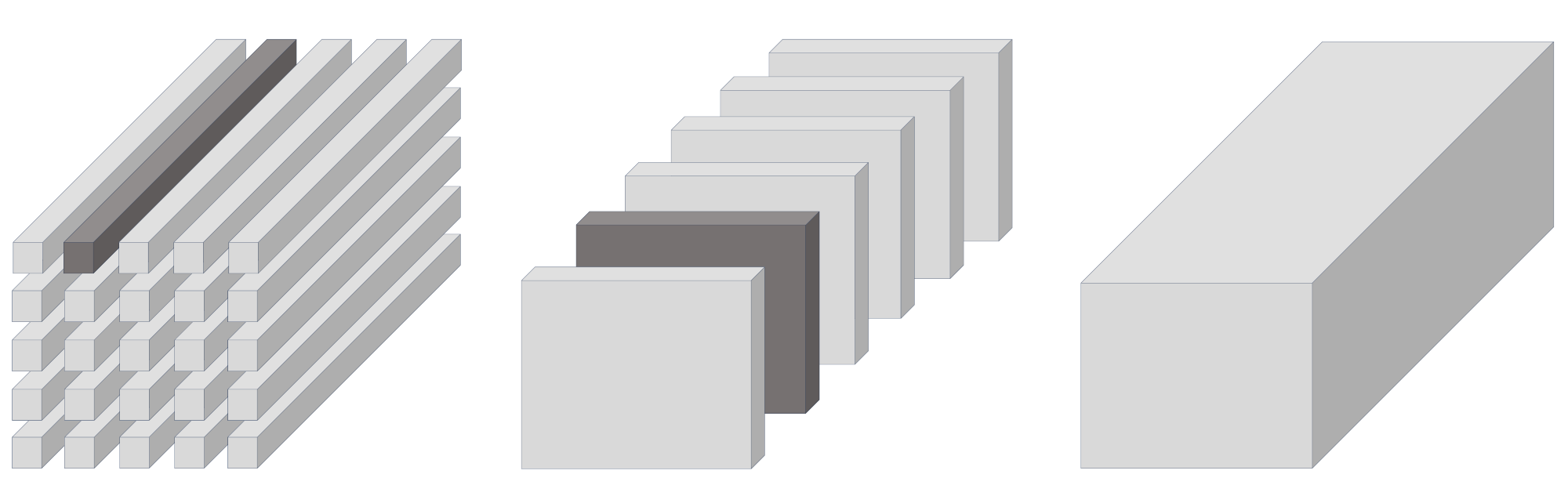}\\
(a) $\mbOmega_{\cdot 12} \in \mathbb{R}^{H}$~~~~~~~~~~~~~ (b) $\mbOmega_{2\cdot\cdot} = \Omega_{(2)} \in \mathbb{R}^{p \times p}$ ~~~~~~~~~~~~ (c) $\mbOmega \in \mathbb{R}^{H \times p \times p}$~~~~~~~~~~~
\end{center}
\caption{Visualization of (a) the fibers of $\mbOmega $ which are penalized
by the final term in \protect \eqref{eq:covEstimator}, (b) the organization of
$\mbOmega $ into the $\Omega _{(h)}$, and (c) the three way tensor
$\mbOmega $.}%
\label{fig:tensorDiagram}
\end{figure}

To describe our estimator, define
$\mbOmega \in \mathbb{R}^{H \times p \times p}$ as the three-way tensor
where
$\mbOmega _{h\cdot \cdot} = \Omega _{(h)} \in \mathbb{R}^{p \times p}$
for $h \in [H]$ and
$\mbOmega _{\cdot jk} = (\Omega _{(1)jk}, \dots , \Omega _{(H)jk})^{
\top }\in \mathbb{R}^{H}$ for $(j,k) \in [p] \times [p]$. We present a
visualization of the tensor $\mbOmega $ in Figure~\ref{fig:tensorDiagram}. The mode-1 fibers, $\mbOmega _{\cdot jk}$, are
vectors containing the $(j,k)$th entry of all the $\Omega _{(h)}$. Assuming
sparsity patterns are shared across populations is thus equivalent to assuming
$\mbOmega ^{*}_{\cdot jk} = 0$ for many pairs $(j,k)$. Finally, let
$\omega _{(h)} = {\mathrm{diag}}(\Omega _{(h)})$ for $h \in [H]$.

Generalizing~\eqref{eq:L1_estimator} with an additional positive definiteness
constraint, we propose to estimate $\mbOmega ^{*}$ using
%
\begin{align}
\label{eq:covEstimator}
&\argmin _{\mbOmega \in \mathbb{R}^{H \times p \times p}} \left\{
\sum _{h=1}^{H} \left (\| \widehat{\Theta}_{(h)} - \omega _{(h)}
\mathds{1}_{p}^{\top }- \mathds{1}_{p}\omega _{(h)}^{\top }+ \hspace{-1pt}2
\Omega _{(h)} \|_{F}^{2} + \lambda \|\Omega _{(h)}^{-}\|_{1} \hspace{-1pt}\right )\hspace{-1pt}+
\gamma \hspace{-1pt}\sum _{j\neq k} \|\mbOmega _{\cdot jk}\|_{2}  \right\}
\\
& ~~~~~~~~~~~~~~~~~~~\text{subject to } ~\Omega _{(h)} = \Omega _{(h)}^{\top},~
\Omega _{(h)} \succcurlyeq \epsilon I_{p} ~ \text{ for all } ~h \in [H], \notag
\end{align}
where $\lambda \geq 0$, $\gamma \geq 0$, and $\epsilon \geq 0$ are user-specified
tuning parameters, $\|\cdot \|_{2}$ denotes the Euclidean norm of a vector,
and $A \succcurlyeq \epsilon I_{p}$ means that $A - \epsilon I_{p}$ is
positive semidefinite.

The estimator~\eqref{eq:covEstimator} imposes both a lasso-type penalty
on the off-diagonal entries of the $\Omega _{(h)}$, as well as a group
lasso penalty on the mode-1 fibers of the tensor $\mbOmega $. Note that
if $H = 1$, taking either $\lambda = 0$ or $\gamma = 0$ with the other
nonzero,~\eqref{eq:covEstimator} simplifies to~\eqref{eq:L1_estimator} with a positive definiteness constraint. For example,
if $\gamma = 0$, then~\eqref{eq:covEstimator} simplifies to the estimator
%
\begin{align}
\argmin _{\Omega _{(h)} \in \mathbb{R}^{p \times p}}& \left ( \|
\widehat{\Theta}_{(h)} - \omega _{(h)} \mathds{1}_{p}^{\top }-
\mathds{1}_{p}\omega _{(h)}^{\top }+ 2 \Omega _{(h)} \|_{F}^{2} +
\lambda \|\Omega _{(h)}^{-} \|_{1} \right ),
\label{eq:SingleCovEstimator}
\\
& ~~~~~~\text{subject to }~\Omega _{(h)} = \Omega _{(h)}^{\top}, ~~
\Omega _{(h)} \succcurlyeq \epsilon I_{p},
\notag
\end{align}
applied to each of the $H$ populations separately. The estimator~\eqref{eq:SingleCovEstimator} can be seen as a convex approximation to~\eqref{eq:L0_estimator} where we have replaced the $L_{0}$ constraint with
an $L_{1}$ constraint, and replaced the feasible set with the closed convex
set
$\{\Omega \in \mathbb{R}^{p \times p}: \Omega = \Omega ^{\top},
\Omega \succcurlyeq \epsilon I\}$. The tuning parameter $\epsilon $ serves
as a lower bound on the smallest eigenvalue of the solution. For this reason,
we do not recommend tuning $\epsilon $, but rather fixing it at some reasonably
small quantity like $10^{-4}$, as in \citet{xue2012positive}.

By taking $\gamma > 0$, however,~\eqref{eq:covEstimator} ties the estimators
of $\Omega _{(1)}^{*}, \dots , \Omega _{(H)}^{*}$ together. For large values
of the tuning parameter $\gamma $, the second penalty in~\eqref{eq:covEstimator} will require that the solution to~\eqref{eq:covEstimator} has some $\mbOmega _{\cdot jk} = 0$, i.e., that
sparsity is partially shared across all $H$ basis covariance matrix estimates.
In the leftmost subfigure of Figure~\ref{fig:tensorDiagram}, (a), we provide
an example of the group of parameters---the (1,2)th element of each
$\Omega _{(h)}$---which are jointly penalized by the group lasso penalty.
The tuning parameter $\gamma $ controls whether this group is entirely
zero or not, whereas the tuning parameter $\lambda $ controls sparsity
in the individual entries of the $\Omega _{(h)}$, as displayed in subfigure
(b).

Importantly,~\eqref{eq:covEstimator} is a convex optimization problem and
as we discuss in Section \ref{sec3}, can be solved using first-order methods.

\subsection{Positive definiteness}
\label{subsec:PD}

To understand why enforcing positive definiteness can be necessary, consider
estimating a single $\Omega ^{*}$ whose off-diagonal entries are assumed
to be zero. Then the problem reduces to estimation of the variances of
the log abundances and $\eqref{eq:L0_estimator}$ admits a closed-form solution.

\begin{prop}
\label{prop1}
If $p \geq 3$, the solution to~\eqref{eq:L0_estimator} with $s = 0$ (or
equivalently,~\eqref{eq:L1_estimator} with $\lambda = \infty $) is unique
and is given by
\begin{equation*}
\widehat{\omega}_{j} = \frac{1}{p - 1}\sum _{k \neq j}
\widehat{\Theta}_{jk} - \frac{1}{2(p - 1)(p - 2)}\sum _{k \neq j}
\sum _{l\neq j} \widehat{\Theta}_{lk}, ~~~ j \in [p].
\end{equation*}
\end{prop}
The factor 2 in the denominator is due to the double sum running over both
the upper and lower triangular parts of the symmetric
$\widehat{\Theta}$. The proposition reveals variance estimates can be negative
if positive definiteness is not enforced. Roughly speaking, for large
$p$, $\widehat{\omega}_{j}$ will be negative if the average of the elements
in $\widehat{\Theta}$ not in the $j$th row or column is larger than the
average of the elements in the $j$th row and column. It is not difficult
to produce such examples. As an illustration, the following
$\widehat{\Theta}$ resulted from simulating $n = 10$ compositional
$x_{i} \in \mathbb{C}^{2}$ by drawing the $\log (W)$ independently
from a multivariate normal distribution with mean zero and identity covariance
matrix:
\begin{equation*}
\widehat{\Theta} =
\begin{pmatrix}
0 & 3.83 & 2.45
\\
3.83 & 0 & 1.24
\\
2.45 & 1.24 & 0
\end{pmatrix}
.
\end{equation*}
Thus, $\widehat{\omega}_{3} = (2.45 + 1.24) / 2 - 3.83 / 2 = -0.07$. Intuitively,
negative variance estimates are more likely when $p$ is large relative
to $n$.

\subsection{Existing estimators}
\label{subsec:existing_estimators}

An alternative estimator of a single basis covariance matrix
$\Omega ^{*}$ is based on the centered log-ratio covariance matrix
\citep{aitchison2003statistical}, $\Gamma ^{*}$, whose $(j,k)$th entry
is
\begin{equation*}
\Gamma ^{*}_{jk} = {\mathrm{Cov}}\left [\log \{X_{j}/g(X)\}, \log \{X_{k}/g(X)
\}\right ]
\end{equation*}
where $g(X) = (\prod _{i=1}^{p} X_{i})^{1/p}$ is the geometric mean of
$X$. Specifically,
\begin{align*}
\Theta ^{*}_{jk} & = {\mathrm{Var}}\{ \log (X_{j}/X_{k})\}
\\
& = {\mathrm{Var}}[ \log \{X_{j}/g(X)\} - \log \{X_{k}/g(X)\}]
\\
& = {\mathrm{Var}}[\log \{X_{j}/g(X)\}] + {\mathrm{Var}}[\log \{X_{k}/g(X)\}] \\
& \quad - 2
{\mathrm{Cov}}[\log \{X_{j}/g(X)\}, \log \{X_{k}/g(X)\}]
\end{align*}
so that
$\Theta ^{*} = \gamma ^{*} \mathds{1}_{p}^{\top }+ \mathds{1}_{p}
\gamma ^{*\top} - 2 \Gamma ^{*}$, where
$\gamma ^{*} = {\mathrm{diag}}(\Gamma ^{*})$. \citet{cao2016large} show there
exists a unique $\Gamma ^{*}$ such that
$\Theta ^{*} = \gamma ^{*} \mathds{1}_{p}^{\top }+ \mathds{1}_{p}
\gamma ^{*\top} - 2 \Gamma ^{*}$ and that
$\max _{j,k}|\Omega ^{*}_{jk} - \Gamma ^{*}_{jk}| \leq (3/p)(\max _{j
\in [p]} \sum _{k=1}^{p}|\Omega ^{*}_{jk}|)$. Thus, by proposing a two-step
procedure to get an estimate of $\Gamma ^{*}$ from
$\widehat{\Theta}$, they also get an indirect estimate of
$\Omega ^{*}$ that can perform well when $p$ is large. However, their estimator
is not guaranteed to be positive definite, nor is it the solution to an
optimization problem amenable to analysis.

\citet{fang2015cclasso} proposed a different estimator, using that with
$F = I_{p} - \mathds{1}_{p} \mathds{1}_{p}^{\top}/p$,
$ F \Omega ^{*} F = F {\mathrm{Cov}}(\log X) F$. Thus, replacing
${\mathrm{Cov}}(\log X)$ with its sample version, say,
$\widehat\Omega _{X}$, a natural estimating equation is
$F(\Omega - \widehat\Omega _{X})F = 0$. To account for differing variances
in each element of $F(\Omega ^{*} - \widehat\Omega _{X})F$, they propose
the weighted least squares estimator
%
\begin{equation}
\label{eq:cclasso}
\argmin _{\Omega = \Omega ^{\top}} \left \{ \frac{1}{2}\|F(\Omega -
\widehat\Omega _{X})F\|_{V}^{2} + \lambda \|\Omega ^{-}\|_{1}\right
\},
\end{equation}
where $V$ is a diagonal matrix with ${\rm diag}(V) = \{{\mathrm{diag}}(F \widehat\Omega _{X} F)\}^{-1}$ and
$\|A\|_{V}^{2} = {\mathrm{tr}}(A^{\top }V A)$. While
\citet{fang2015cclasso} suggest including a positive definiteness constraint
on the optimization variable $\Omega $ in~\eqref{eq:cclasso}, their computational
algorithm does not enforce this constraint. Instead, if the solution to~\eqref{eq:cclasso} is not positive definite, they estimate
$\Omega ^{*}$ using its nearest positive definite matrix. This can be appropriate,
but often leads to a non-sparse estimate \citep{sun2015decomposition}.

In contrast to the methods of \citet{cao2016large} and
\citet{fang2015cclasso}, our estimator is ``direct'' in the sense that
we do not rely on estimation of intermediate quantities like
$\Gamma ^{*}$, nor do we rely on post-hoc adjustments to achieve positive
definiteness.

Many other estimators of $\Omega ^{*}$ exist, though we do not cover them
in detail here. In general, these estimators do not enforce both positive
definiteness and sparsity, and are not specifically designed to estimate
multiple covariance matrices simultaneously: see
\citet{friedman2012inferring,ban2015investigating,he2021robust,li2022robust},
for example, and see \citet{ma2021networks} for a comprehensive review.

As mentioned in Section~\ref{sec:Introduction}, our work is related to
the literature on jointly estimating sparse precision (inverse covariance)
matrices for multiple populations
\citep[e.g.,][]{guo2011joint,danaher2014joint,price2015ridge,saegusa2016joint, price2021estimating}.
However, our work is distinct in at least two important ways. First, these
existing methods are largely focused on estimating precision matrices,
rather than covariance matrices. Sparse precision matrix estimation and
sparse covariance matrix estimation are fundamentally different tasks.
Second, these methods, broadly speaking, assume the observed data are normally
distributed or utilize a normal negative log-likelihood as a loss function.
These methods are thus not directly applicable to either covariance matrix
estimation, nor the analysis of compositional data.

Differences in motivation aside, there are some similarities between our
work and that of \citet{guo2011joint} and \citet{danaher2014joint}, for
example. \citet{danaher2014joint} use the same penalty as~\eqref{eq:covEstimator}, but applied to precision matrices for normally
distributed data. \citet{guo2011joint} use a group lasso-type penalty to
encourage shared sparsity patterns across populations in the same context
as \citet{danaher2014joint}. The existing work most closely related to
that of our own is \citet{bigot2011group}, who use a group lasso penalty
in the context of estimating multiple sparse covariance matrices from data
observed with additive noise. One could not straightforwardly use their
method for estimating basis covariance matrices from compositional data,
and moreover, neither their theory nor algorithms apply to our estimator.

\section{Computation}
\label{sec3}

\subsection{Proximal-proximal gradient descent algorithm}
\label{sec3.1}

In order to solve the optimization problem to compute~\eqref{eq:covEstimator}, we must address both the nondifferentiability
of the objective function and the positive definiteness constraint. To
do so, we use the proximal-proximal gradient descent algorithm
\citep{Davis2017}, which allows us to handle the nondifferentiable penalty
and positive definiteness constraint separately. The algorithm generalizes
the well-known proximal gradient descent algorithm
\citep[Section 4.2]{parikh2014proximal} to handle problems where the objective
function to be minimized is the sum of three convex functions. Specifically,
supposing $f$ and $g$ are closed, proper, and convex functions; and
$\ell $ is convex and differentiable with $\beta ^{-1}$-Lipschitz continuous
gradient for some $\beta > 0$; consider a problem of the form
%
\begin{align}
\label{eq:DavisEq}
\minim _{u \in \mathbb{R}^d} \left \{\ell (u) + f(u) + g(u) \right \}.
\end{align}
Further suppose there exists
$u^{\star }\in \mathbb{R}^{d}$ such that
$0 \in \partial f(u^{\star}) + \partial g(u^{\star}) + \nabla \ell (u^{
\star})$ where $\partial f(u)$ denotes the subdifferential of $f$ at
$u$. The proximal operator of a function $f$ evaluated at $u$ is
\begin{equation*}
\textbf{prox}_{f}(u) = \argmin _{y \in {\mathrm{dom}} f} \left \{
\frac{1}{2}\|u - y\|_{2}^{2} + f(y) \right \}.
\end{equation*}
\citet{Davis2017} show that~\eqref{eq:DavisEq} can be solved by an algorithm
whose $(t)$th iterates are computed using the updating equations
\begin{align*}
u_{g}^{(t)} &= \textbf{prox}_{\alpha g}(v^{(t)})
\\
u_{f}^{(t)} &= \textbf{prox}_{\alpha f}\{ 2u_{g}^{(t)} - v^{(t)} -
\alpha \nabla \ell (u_{g}^{(t)}) \}
\\
v^{(t+1)} &= v^{(t)} + u_{f}^{(t)} - u_{g}^{(t)},
\notag
\end{align*}
where $v^{(0)}$ is an arbitrary point in $\mathbb{R}^d$ and
$\alpha \in (0, 2\beta )$ is fixed. Here, the superscript $(t)$ denotes
the $(t)$th iterate. As $t \to \infty $, $u_{g}^{(t)} \to u^{\star}$ and
$u_{f}^{(t)} \to u^{\star}$ \citep{Davis2017}. In practice, however, this
algorithm can be slow to converge: fixing the step size
$\alpha \in (0, 2\beta )$ can sometimes lead to incremental progress. Therefore,
we use a modified version of the proximal-proximal gradient descent algorithm
proposed by \citet{pedregosa18}, which allows us to start with a step size $\alpha $ larger than $2\beta $ and reduce its
value as needed, thus potentially accelerating the descent. The $(t+1)$th iterates
of the algorithm use the updating equations
%
\begin{align}
u_{f}^{(t+1)} &= \textbf{prox}_{\alpha f}\{ u_{g}^{(t)} - \alpha v^{(t)}
- \alpha \nabla \ell ( u_{g}^{(t)}) \}
\label{eq:firstApd}
\\
u_{g}^{(t+1)} &= \textbf{prox}_{\alpha g}(u_{f}^{(t+1)} + \alpha v^{(t)})
\label{eq:secondApd}
\\
v^{(t+1)} &= v^{(t)} + \alpha ^{-1}( u_{f}^{(t+1)} - u_{g}^{(t+1)}).
\label{eq:thirdApd}
\end{align}
At each step, after~\eqref{eq:firstApd} is carried out, the value
\begin{equation*}
Q(u_{f}^{(t+1)}, \alpha ) = \ell (u_{g}^{(t)}) + \langle \nabla
\ell (u_{g}^{(t)}), u_{f}^{(t+1)} - u_{g}^{(t)} \rangle +
\frac{1}{2\alpha} \| u_{f}^{(t+1)} - u_{g}^{(t)}\|_2^2
\end{equation*}
\noindent
is compared to $\ell (u_{f}^{(t+1)})$. If
$\ell (u_{f}^{(t+1)}) \leq Q(u_{f}^{(t+1)}, \alpha )$, then the algorithm
proceeds to~\eqref{eq:secondApd}. If
$\ell (u_{f}^{(t+1)}) > Q(u_{f}^{(t+1)}, \alpha )$, then $\alpha $ is replaced
with $\tau \alpha $, where $\tau \in (0,1)$ is a constant, and~\eqref{eq:firstApd} is carried out again. This process is repeated until
$\ell (u_{f}^{(t+1)}) \leq Q(u_{f}^{(t+1)}, \alpha )$.

The efficiency of this algorithm hinges on the ability to compute the proximal
operators of the functions $g$ and $f$ efficiently. As we will show momentarily,
we can write the optimization problem from~\eqref{eq:covEstimator} as~\eqref{eq:DavisEq} and the corresponding $g$ and $f$ have proximal operators
which can be solved in closed form.

\subsection{Application to proposed estimator}
\label{sec3.2}

In order to express the problem in~\eqref{eq:covEstimator} in a form analogous
to~\eqref{eq:DavisEq}, we must define the corresponding $\ell $, $f$, and
$g$. First, let
$\chi _{\epsilon}: \mathbb{R}^{p \times p} \to \{0, \infty \}$ be the function
$\chi _{\epsilon}(\Omega ) = \infty \cdot \mathbf{1}(\{\epsilon I_{p}
\succ \Omega \} \cup \{\Omega \neq \Omega ^{\top}\})$, with the convention
$\infty \cdot 0 = 0$. Then, the unconstrained objective function from~\eqref{eq:covEstimator} is
%
\begin{equation}
\begin{split}
\label{eq:reformulation}
\sum _{h=1}^{H} \bigl  \{\| \widehat{\Theta}_{(h)} - \omega _{(h)}
\mathds{1}_{p}^{\top }- \mathds{1}_{p}\omega _{(h)}^{\top }+ 2
\Omega _{(h)} \|_{F}^{2} + \lambda \|\Omega _{(h)}^{-}\|_{1} &+ \chi _{
\epsilon}(\Omega _{(h)}) \bigr \} + \gamma \sum _{j \neq k} \|
\mbOmega _{\cdot jk}\|_{2}.
\end{split}
\end{equation}
If we minimize~\eqref{eq:reformulation} over all
$\mbOmega \in \mathbb{R}^{H \times p \times p}$, the minimizer with respect
to each $\Omega _{(h)}$ must belong to the set
$\{\Omega \in \mathbb{R}^{p \times p}: \Omega = \Omega ^{\top},
\Omega \succcurlyeq \epsilon I_{p} \}$. Thus, defining
$\ell (\mbOmega ) = \sum _{h=1}^{H} \| \widehat{\Theta}_{(h)} -
\omega _{(h)} \mathds{1}^{\top }- \mathds{1}\omega _{(h)}^{\top }+ 2
\Omega _{(h)} \|_{F}^{2}$,
$f(\mbOmega ) = \lambda \sum _{h=1}^{H} \|\Omega _{(h)}^{-}\|_{1} +
\gamma \sum _{j \neq k} \|\mbOmega _{\cdot jk}\|_{2}$, and
$g(\mbOmega ) = \sum _{h=1}^{H} \chi _{\epsilon}(\Omega _{(h)})$,~\eqref{eq:reformulation} has the form of~\eqref{eq:DavisEq}. Moreover,
$f$ and $g$ are closed, proper, and convex functions; and the function
$\ell $ is convex and differentiable with Lipschitz continuous gradient.

Specifically, letting
$\widehat{\mbTheta} \in \mathbb{R}^{H \times p \times p }$ be the three-way
tensor made up of
$\widehat{\Theta}_{(1)}, \dots ,\\\widehat{\Theta}_{(H)}$ so that
$\widehat{\mbTheta}_{hjk} = \widehat{\Theta}_{(h)jk}$, the function
$\ell $ is differentiable with respect to $\mbOmega $ with gradient
\begin{align*}
\left [\nabla \ell (\mbOmega ) \right ]_{hjk} =
\begin{cases}
\sum \limits _{l \in [p]\setminus \{j\}} (4\mbOmega _{hjj} - 4
\widehat{\mbTheta}_{hjl} -8 \mbOmega _{hjl} + 4\mbOmega _{hll}) &:j = k
\\
8\mbOmega _{hjk} - 4\mbOmega _{hjj} - 4\mbOmega _{hkk} + 4
\widehat{\mbTheta}_{hjk} &:j \neq k
\end{cases},
\end{align*}
for all $(h,j,k) \in [H] \times [p] \times [p]$. The updating equations
corresponding to~\eqref{eq:firstApd}--\eqref{eq:thirdApd} are
%
\begin{align}
\bovtex{\Omega}^{(t+1)} & = \argmin _{\bovtex{\Omega} \in \mathbb{R}^{H
\times p \times p}} \bigg\{\frac{1}{2}
\vertiii{\bovtex{\Omega} - \bovtex{\Lambda}^{(t)}}_{F}^{2} + \alpha \lambda
\sum _{h=1}^{H} \|\mbOmega _{(h)}^{-}\|_{1} + \alpha \gamma \sum _{j
\neq k} \|\bovtex{\Omega}_{\cdot jk}\|_{2} \bigg\}
\label{eq:OmegaUp}
\\
\bovtex{\tilde{\Omega}}^{(t+1)} &= \argmin _{\bovtex{\Omega} \in \mathbb{R}^{H
\times p \times p}} \bigg\{\frac{1}{2}
\vertiii{\bovtex{\Omega} - \bovtex{\Omega}^{(t+1)} - \alpha \bovtex{\Psi}^{(t)}}_{F}^{2}
+ \alpha \sum _{h=1}^{H} \chi _{\epsilon}(\Omega _{(h)})\bigg\}
\label{eq:tildeOmegaUp}
\\
\bovtex{\Psi}^{(t+1)} &= \bovtex{\Psi}^{(t)} + \alpha ^{-1}(\bovtex{\Omega}^{(t+1)}
- \bovtex{\tilde{\Omega}}^{(t+1)} ),
\label{eq:ThetaUp}
\end{align}
where
$\bovtex{\Lambda}^{(t)} = \bovtex{\tilde{\Omega}}^{(t)} - \alpha \bovtex{\Psi}^{(t)}
- \alpha \nabla \ell (\bovtex{\tilde{\Omega}}^{(t)} )$ and
$\vertiii{\bovtex{A}}_{F}^{2} = \sum _{h,j,k} \bovtex{A}_{hjk}^{2}$ for a three-way
tensor $\bovtex{A}$. Because~\eqref{eq:ThetaUp} is immediate, we focus on~\eqref{eq:OmegaUp} and~\eqref{eq:tildeOmegaUp}.

First,~\eqref{eq:OmegaUp} can be separated across the second and third
mode of $\bovtex{\Omega}$ since for all $(j,k) \in [p] \times [p]$ such that
$j \neq k$,
%
\begin{align}
\label{eq:sparse_group_lasso}
\bovtex{\Omega}_{\cdot jk }^{(t+1)} = \argmin _{\nu \in \mathbb{R}^{H}}
\left \{\frac{1}{2} \| \nu - \bovtex{\Lambda}^{(t)}_{\cdot jk} \|_{2}^{2} +
{\alpha \lambda} \left \|\nu \right \|_{1} + \alpha \gamma \left \|
\nu \right \|_{2} \right \},
\end{align}
and
$\bovtex{\Omega}_{\cdot jj}^{(t+1)} = \bovtex{\Lambda}^{(t)}_{\cdot j j}$ for
$j \in [p]$. The solution to~\eqref{eq:sparse_group_lasso} is
\begin{align*}
\bovtex{\Omega}_{\cdot jk}^{(t+1)} = \left (1 -
\frac{\alpha \gamma}{\|\textbf{soft}(\bovtex{\Lambda}^{(t)}_{\cdot jk}, \alpha \lambda ) \|_{2}}
\right )_{+}\textbf{soft}\left (\bovtex{\Lambda}^{(t)}_{\cdot jk}, \alpha
\lambda \right ),
\end{align*}
where $(a)_{+} = \max (a, 0)$ and
$\textbf{soft}(y, \tau ) = \max (|y| - \tau , 0){\mathrm{sign}}(y)$ is applied
elementwise \citep{SGL_paper}. The second step,~\eqref{eq:tildeOmegaUp}, also has a closed form solution. In particular,~\eqref{eq:tildeOmegaUp} can be solved with respect to each
$\Omega _{(h)}$ separately, in parallel, using that
\begin{align*}
\Omega _{(h)}^{(t+1)} & = \argmin _{\Omega _{(h)} \in \mathbb{R}^{p
\times p}} \left \{ \frac{1}{2} \|\Omega _{(h)} - \Omega _{(h)}^{(t+1)}
- \alpha \boldsymbol{\Psi}_{h \cdot \cdot}^{(t)}\|_{F}^{2} + \chi _{
\epsilon} \left (\Omega _{(h)}\right )\right \}
\\
& = \sum _{j=1}^{p} u_{(h)j} u_{(h)j}^{\top }\max (\xi _{(h)j},
\epsilon ),
\end{align*}
where $u_{(h)j}$ and $\xi _{(h)j}$ are the $j$th eigenvector and eigenvalue
of
$\Omega _{(h)}^{(t+1)} + \alpha {\boldsymbol{\Psi}}_{h \cdot \cdot}^{(t)}$,
respectively, for $h \in [H]$ \citep{Henrion2012}. This is the projection
of
$\Omega _{(h)}^{(t+1)} + \alpha \boldsymbol{\Psi}_{h \cdot \cdot}^{(t)}$
onto the convex set
$\{\Omega \in \mathbb{R}^{p \times p}: \Omega \succcurlyeq \epsilon I_{p}
\text{ and } \Omega = \Omega ^{\top}\}$.

The convergence of the algorithm follows immediately from results in
\citet{pedregosa18}. The specific algorithm we implement is given in Algorithm~\ref{alg1}. Without the positive definiteness constraint (e.g., by taking
$\epsilon = -\infty $), a version of this algorithm simplifies to the standard
proximal gradient descent algorithm
\citep[Chapter 4.2]{parikh2014proximal}.

Enforcing the positive definiteness constraint on each of the
$\Omega _{(h)}$ requires the eigendecomposition of a $p \times p$ matrix, a computation costing $O(p^{3})$ floating point operations. To reduce
the computational burden imposed by this additional constraint, our software
implementation first solves~\eqref{eq:covEstimator} without the positive
definiteness constraint. To do so, we use accelerated proximal gradient
descent \citep[Section 4.3]{parikh2014proximal}. If the solution to the
unconstrained problem satisfies the constraint, then we know this is also
the solution to the constrained problem. If the solution does not satisfy
the constraint, we then apply the proximal-proximal gradient descent algorithm,
Algorithm~\ref{alg1}, initializing at the solution to the unconstrained
problem. This scheme can significantly improve the computing time, especially
when $n$ is large relative to $p$.

An R package implementing our estimators, along with code for reproducing
the results in Section~\ref{sec:Simulations}, can be downloaded from
\texttt{https://github.com/ajmolstad/SpPDCC}.

\begin{algorithm}[t]%
\caption{Adaptive proximal-proximal gradient descent algorithm for computing multiple covariance matrices for compositional data.}%
\label{alg1}
    Initialize $\bovtex{\Psi}^{(0)} \in \mathbb{R}^{H \times p \times p}, \bovtex{\tilde{\Omega}}^{(0)} \in \mathbb{R}^{H \times p \times p}$,  $\alpha > 0$, and $\tau \in (0,1)$. Set $t = 0$ and proceed to \textbf{1.}
    %
\begin{itemize}
    \item[]\textbf{1.} For $(j,k) \in [p] \times  [p]$
\begin{itemize}
            \item[] \textbf{1.1.} If $j = k$
\begin{itemize}
                \item[] \textbf{1.1.1.} Set $\bovtex{\Omega}_{\cdot jj}^{(t+1)} = \bovtex{\tilde{\Omega}}^{(t)}_{\cdot jj} - \alpha \bovtex{\Psi}_{\cdot jj}^{(t)} - \alpha [\nabla \ell (\bovtex{\tilde{\Omega}}^{(t)})]_{\cdot jj}$
            \end{itemize}
\item[] \textbf{1.2.} If $j \neq k$
\begin{itemize}
                \item[] \textbf{1.2.1.} Set $y = \bovtex{\tilde{\Omega}}^{(t)}_{\cdot jk } - \alpha \bovtex{\Psi}_{\cdot jk }^{(t)} - \alpha [\nabla \ell (\bovtex{\tilde{\Omega}}^{(t)})]_{\cdot jk }$
                \item[] \textbf{1.2.2.} Set $w_{h} =(|y_{h}| - \alpha \lambda )_{+}\text{sign}(y_{h})$ for $h \in [H]$
                \item[] \textbf{1.2.3.} Set $\bovtex{\Omega}_{\cdot jk }^{(t+1)} = \left (1 - \frac{\alpha \gamma}{\|w \|_{2}}\right )_{+}w$
            \end{itemize}
\end{itemize}
\item[] \textbf{2.} If $\ell (\bovtex{\Omega}^{(t+1)}) \leq Q(\bovtex{\Omega}^{(t+1)}, \alpha )$, proceed to \textbf{3.} Else, set $\alpha = \tau \alpha $, and return to \textbf{1.}\smallskip
    \item[]\textbf{3.} For $h \in [H]$
\begin{itemize}
        \item[]\textbf{3.1.} Decompose $\bovtex{\Omega}_{h\cdot \cdot}^{(t+1)} + \alpha \bovtex{\Psi}_{h\cdot \cdot}^{(t)} = \sum _{j=1}^{p} \xi _{j} \bovtex{u}_{j} \bovtex{u}_{j}^{\top}$, where $\bovtex{u}_{j}^{\top }\bovtex{u}_{k} = 0$ for $j \neq k$ and $\|\bovtex{u}_{j}\|_{2} = 1$ for all $j \in [p]$
        \item[]\textbf{3.2.} Set $\bovtex{\tilde{\Omega}}_{h\cdot \cdot}^{(t+1)} = \sum _{j=1}^{p} \text{max}(\xi _{j}, \epsilon )\bovtex{u}_{j}\bovtex{u}_{j}^{\top}$
    \end{itemize}

   \item[]\textbf{4.} Set $\bovtex{\Psi}^{(t+1)} = \bovtex{\Psi}^{(t)} + \alpha ^{-1}(\bovtex{\Omega}^{(t+1)} - \bovtex{\tilde{\Omega}}^{(t+1)})$
   \item[]\textbf{5.} If the objective function value converged, terminate the algorithm. Else, set $t = t+1$ and return to step \textbf{1.}
\end{itemize}

\end{algorithm}

\section{Practical considerations}
\label{eq:practicalSection}

To select tuning parameters $(\lambda , \gamma )$, we use $V$-fold cross-validation.
Given candidate tuning parameter sets $\boldsymbol{\lambda}$ and
$\boldsymbol{\gamma}$, for~\eqref{eq:covEstimator} we select tuning parameters
according to
\begin{equation*}
\argmin _{(\lambda , \gamma ) \in \boldsymbol{\lambda}\times
\boldsymbol{\gamma}} \sum _{v=1}^{V} \sum _{h=1}^{H} \|
\widehat{\Theta}_{(h),v} - \tilde{\omega}_{(h),-v}^{\lambda , \gamma}
\mathds{1}_{p}^{\top }- \mathds{1}_{p} [\tilde{\omega}_{(h),-v}^{
\lambda , \gamma}]^{\top }+ 2
\hspace{2pt}
\widetilde{\Omega}_{(h),-v}^{\lambda , \gamma}\|_{F}^{2},
\end{equation*}
where $\widetilde{\mbOmega}_{-v}^{\lambda , \gamma}$ is the solution to~\eqref{eq:covEstimator} with input sample variation matrices
$\widehat{\mbTheta}_{-v}$, which are computed using all the data from outside
the $v$th fold.

In some applications, it may be preferable to let populations with larger
samples sizes have a greater effect on the objective function. To do so,
we propose an alternative variation of~\eqref{eq:covEstimator}, defined as the argument minimizing
%
\begin{align}
\label{eq:covEstimator_weighted}
\sum _{h=1}^{H} \left (\frac{n_{(h)}}{N}\| \widehat{\Theta}_{(h)} -
\omega _{(h)} \mathds{1}_{p}^{\top }- \mathds{1}_{p}\omega _{(h)}^{
\top }+ 2 \Omega _{(h)} \|_{F}^{2} + \lambda \|\Omega _{(h)}^{-}\|_{1}
\right ) + \gamma \sum _{j\neq k} \|\mbOmega _{\cdot jk}\|_{2}
\end{align}
%
%
subject to $\Omega _{(h)} = \Omega _{(h)}^{\top},~ \Omega _{(h)}
\succcurlyeq \epsilon I_{p} ~~ \text{ for all } ~h \in [H],$ where $N = \sum _{h=1}^{H} n_{(h)}$. The objective function \eqref{eq:covEstimator_weighted} accounts for the distinct sample sizes
by weighting each populations' contribution to the likelihood according
to its relative contribution to the total sample size $N$. When using
this estimator, we also recommend modifying the cross-validation criterion
so that the tuning parameter pair selected is
\begin{equation*}
\argmin _{(\lambda , \gamma ) \in \boldsymbol{\lambda}\times
\boldsymbol{\gamma}} \sum _{v=1}^{V} \sum _{h=1}^{H}
\frac{n_{(h),v}}{N_{v}}\|\widehat{\Theta}_{(h),v} - \tilde{\omega}_{(h),-v}^{
\lambda , \gamma} \mathds{1}_{p}^{\top }- \mathds{1}_{p} [
\tilde{\omega}_{(h),-v}^{\lambda , \gamma}]^{\top }+ 2
\hspace{2pt}
\widetilde{\Omega}_{(h),-v}^{\lambda , \gamma}\|_{F}^{2},
\end{equation*}
where $n_{(h),v}$ is the number of samples in the $v$th fold from the
$h$th population, and $N_{v} = \sum _{h=1}^{H} n_{(h),v}$. We compare the
performance of~\eqref{eq:covEstimator_weighted} to~\eqref{eq:covEstimator}, among other competitors, in the Appendix.

\section{Statistical properties}
\label{sec5}

\subsection{Asymptotics for single population estimator}
\label{sec5.1}

Though our primary focus is the multipopulation estimator~\eqref{eq:covEstimator}, the estimator~\eqref{eq:SingleCovEstimator} is
itself a novel and useful estimator of a single basis covariance matrix.
In this section, we study its asymptotic properties. Specifically, we study
$\widehat\Omega $ defined as
%
\begin{equation}
\label{eq:L1_estimator_theory}
\argmin _{\Omega = \Omega ^{\top}} \left \{ \|\widehat{\Theta} -
\omega \mathds{1}_{p}^{\top }- \mathds{1}_{p} \omega ^{\top }+ 2
\Omega \|_{F}^{2} + \lambda \|\Omega ^{-}\|_{1}\right \}~~
\text{subject to}~~ \Omega \succcurlyeq \epsilon I_{p}.
\end{equation}

We will require the following assumptions.
\begin{itemize}
\item[] \textbf{A1.} (Sub-Gaussian log abundances). The sample variation
matrix, $\widehat{\Theta}$, is computed from $n$ independent and identically
distributed samples $W = (W_{1}, \dots , W_{p})^{\top}$ such that each
$\log (W_{j})$ is sub-Gaussian and
${\mathrm{Cov}}\{\log (W)\} = \Omega ^{*}$.
\smallskip
\item[] \textbf{A2.} (Row-wise sparsity). As $n \to \infty $,
$\max _{j} s_{j} / p \to 0$ where $s_{j}$ is the number of nonzero off-diagonal
entries in the $j$th row of $\Omega ^{*}$.
\smallskip
\item[] \textbf{A3.} (Alignment of $n$ and $p$). As $n \to \infty $,
$p \to \infty $ and $\log (p) / n \to 0$.
\end{itemize}
The assumptions \textbf{A1}---\textbf{A3} are natural in the context of high-dimensional
compositional data. Assumption \textbf{A2} is needed to establish the restricted
strong convexity of the loss function
$\|\widehat{\Theta} - \omega \mathds{1}_{p}^{\top }- \mathds{1}_{p}
\omega ^{\top }+ 2 \Omega \|_{F}^{2}$ in a neighborhood of
$\Omega ^{*}$. \citet{cao2016large} assume similar sparsity, which in their
setting ensures approximate identifiability (see their Proposition 1 and
the comments following it). An analogous interpretation is possible here:
unidentifiable parameters often lead to loss functions that are constant
in some directions, but Assumption \textbf{A2} ensures the loss function is
strictly convex around $\Omega ^{*}$ in the directions that matter
\citep[see][Section 2.4, for details]{Negahban.etal2012}. For this assumption
to hold, we need $p$ to grow with $n$. This is congruous with the assumptions
in \citet{cao2016large}, who require $p \to \infty $ as
$n \to \infty $ for consistency and characterize this as a ``blessing of
dimensionality''. Assumption \textbf{A3} is standard in high-dimensional
covariance matrix estimation.

We now state our first result concerning the asymptotic error of our estimator.
The proof of this and all subsequent results can be found in the Appendix.
Recall $ s= \sum _{j=1}^{p} s_{j}$ and let $\varphi _{p}$ be the $p$th
largest eigenvalue of its matrix-valued argument.
%
\begin{thm}%
\label{thm:singlePop_error}
Suppose \textbf{A1}--\textbf{A3} hold. If
$\epsilon < \varphi _{p}(\Omega ^{*})$ and
$\lambda = \sqrt{c_{1} \log (p) /n}$ for fixed constant $c_{1} > 0$ sufficiently
large, then there exists a constant $b_{1} \in (0, \infty )$ such that
%
\begin{equation}
\label{bound:OnePop}
\frac{\|[\widehat\Omega - \Omega ^{*}]^{-}\|_{F}}{\sqrt{p}} + \|
\widehat\omega - \omega ^{*} \|_{2} \leq b_{1} \left (\sqrt{
\frac{s \log (p)}{pn} }+ \sqrt{\frac{p\log (p)}{n}}\right )%
\end{equation}
and
$\|\widehat\omega - \omega ^{*} \|_{2} \leq b_{1} \sqrt{p \log (p)/n}$
with probability tending to one as $n \to \infty $.
\end{thm}
The error bound in~\eqref{bound:OnePop} consists of two parts: the error
for estimating off-diagonals and the diagonals. The asymptotic Euclidean
norm error for the diagonals is $b_{1}\sqrt{p \log (p)/n}$. The Frobenius
norm error for the off-diagonals, however, cannot be disentangled from
the diagonal error. Though our results would seem to suggest that
$\|[\widehat\Omega - \Omega ^{*}]^{-}\|_{F} \leq b_{1} \sqrt{s \log (p)
/n}$ with probability tending to one, we are only able to establish a bound for
$\|[\widehat\Omega - \Omega ^{*}]^{-}\|_{F}/\sqrt{p} + \|
\widehat\omega - \omega ^{*} \|_{2} $. We cannot isolate the asymptotic
error for the off-diagonals because of the intrinsic connection between
the diagonals and off-diagonals in the objective function~\eqref{eq:L1_estimator_theory}. This is in contrast to some traditional
covariance matrix estimators, where off-diagonals can be estimated in a
way which is not dependent on the diagonals.

Note that although~\eqref{eq:L1_estimator_theory} is the solution to a
penalized least squares problem, we do not assume any type of restricted
eigenvalue condition \citep{raskutti2010restricted}. Instead, in our proof
we first show that $\widehat\Omega - \Omega ^{*}$ belongs to a restricted
set, then establish a quadratic lower bound on
$\ell (\widehat\Omega ) - \ell (\Omega ^{*}) - {\mathrm{tr}}\{\nabla \ell (
\Omega ^{*})^{\top }(\widehat\Omega - \Omega ^{*})\}$ over this set where
here, $\ell $ is the objective function from~\eqref{eq:L1_estimator} with
$\lambda = 0$. Our technique for establishing this bound may be applicable
in other penalized least squares problems.

Direct comparison of our estimation error bound to those established in
\citet{cao2016large} is not possible as their results are given in terms
of the spectral norm, and under a different set of assumptions.

\subsection{Asymptotics for multiple population estimator}
\label{sec5.2}

Next, we consider the multiple population estimator~\eqref{eq:covEstimator} with $\lambda = 0$. By doing so, we are able to
illustrate how our method exploits shared sparsity across the populations.
Our results will apply with $N = \sum _{h=1}^{H} n_{(h)}$ tending to infinity.
To establish error bounds for this estimator, we will need a slightly different
set of assumptions than in the single population case.
\begin{itemize}
\item[] \textbf{A4.} (Bounded log abundances). The sample variation matrix
$\widehat{\Theta}_{(h)}$ is computed from $n_{(h)}$ independent and identically
distributed samples $W_{(h)} = (W_{(h)1}, \dots , W_{(h)p})^{\top}$ such
that $\log (W_{(h)k}) \in [-L, L]$ for all $k \in [p]$ and
${\mathrm{Cov}}\{\log (W_{(h)})\} = \Omega _{(h)}^{*}$ for all
$h \in [H]$.
\smallskip
\item[] \textbf{A5.} (Fiber-wise sparsity). As $N \to \infty $,
$\max _{j} \tilde{s}_{j}/p \to 0$ where
$\tilde{s}_{j} = |\{k: \mbOmega ^{*}_{\cdot jk} \neq 0, k \neq j\}|$ for
$j \in [p]$.%
\smallskip
\item[] \textbf{A6.} (Nonvanishing $n_{(h)}/N$). There exists
$\pi > 0$ such that for $N$ sufficiently large,
$\min _{h\in [H]} n_{(h)}/N \geq \pi $.
\smallskip
\item[] \textbf{A7.} (Alignment of $n_{(h)}$, $p$, $H$, and $L$). As
$N \to \infty $, $p \to \infty $, $\log (p) / n_{(h)} \to 0$, and
$L^{4} H/ n_{(h)} \to 0$ for all $h \in [H]$.
\end{itemize}
Assumption \textbf{A4} requires that the log abundances take values over
the interval $[-L,L]$. When $W_{(h)k}$ is a normalized count---as is standard
in microbiome data---this assumption requires that all counts are bounded
away from zero and infinity. A positive lower bound on $W_{(h)k}$ is often
assumed implicitly in the analysis of compositional data. Of course,
\textbf{A4} is stronger than \textbf{A1}, but allows us to establish a concentration
inequality on the Euclidean norm of the fibers of
$\nabla \ell (\mbOmega ^{*})$. The quantity $L$ will appear in our asymptotic
error bound, so this assumption is not so restrictive since $L$ can be
arbitrarily large.

Assumption \textbf{A5} requires that the number of nonzero off-diagonal
entries in any of the $\Omega ^{*}_{(h)}$ does not grow too quickly with
$p$. Like \textbf{A2}, \textbf{A5} implicitly requires that $p$ grows as
$N \to \infty $. Assumption \textbf{A6} is a requirement on how frequently,
as $N \to \infty $, we sample from each of the $H$ populations. This assumption
requires that we do not systemically undersample from any of the $H$ populations.
Our error bounds will depend on $\pi $, so we can quantify how sampling
affects estimation accuracy.

We are ready to state our asymptotic error bound for~\eqref{eq:covEstimator} with $\lambda = 0$. Our bound will depend on
$\tilde{s} = \sum _{j=1}^{p} \tilde{s}_{j}$.
%
\begin{thm}%
\label{thm:MultiPop_error}
Suppose \textbf{A4}--\textbf{A7} hold. Define $\widehat{\mbOmega}$ as the
solution to~\eqref{eq:covEstimator} with $\lambda = 0$. Let
$\boldsymbol{\omega}^{*} = (\omega ^{*}_{(1)}, \dots , \omega ^{*}_{(H)})$
and
$\widehat{\mbomega }= (\widehat\omega _{(1)}, \dots , \widehat\omega _{(H)})$.
If $\epsilon < \min _{h \in [H]} \varphi _{p}(\Omega ^{*}_{(h)})$ and
$\gamma = \sqrt{c_{2} L^{4} H / \pi N} + \sqrt{c_{2} \log (p) / \pi N}$
for fixed constant $c_{2} > 0$ sufficiently large, then there exists a
constant $b_{2} \in (0, \infty )$ such that
\begin{equation*}
\frac{\vertiii{[\widehat{\mbOmega} - \mbOmega ^{*}]^{-}}_{F}}{\sqrt{p}}
+ \|\widehat{\mbomega }- \mbomega ^{*}\|_{F} \leq b_{2} \left \{
\left (\frac{\sqrt{\tilde{s}} + p}{\sqrt{p}}\right ) \left (
\sqrt{\frac{L^{4} H}{\pi N}} + \sqrt{\frac{\log (p) }{\pi N }} \right )
\right \}
\end{equation*}
and
\begin{equation*}
\|\widehat{\mbomega }- \mbomega ^{*}\|_{F} \leq b_{2} \left (\sqrt{
\frac{p L^{4} H}{\pi N}} + \sqrt{\frac{p \log (p) }{\pi N }}\right )
\end{equation*}
with probability tending to one as $N \to \infty $.
\end{thm}

The bound in Theorem~\ref{thm:MultiPop_error} can be interpreted in a similar
way as the bound in Theorem~\ref{thm:singlePop_error}. Specifically, we
cannot separate the error for estimating the off-diagonals of the
$\Omega ^{*}_{(h)}$ from the error for estimating the diagonals. In particular,
where the diagonals and off-diagonals affect the error bound are through
their contribution to numerator in the leftmost term of the error bound:
the $\sqrt{\tilde{s}}$ comes from having to estimate nonzero entries in
$\tilde{s}$ off-diagonals of the $\Omega ^{*}_{(h)}$, whereas the
$\sqrt{p}$ comes from having to estimate $p$ diagonal entries in each
$\Omega ^{*}_{(h)}$.

Just as in Theorem~\ref{thm:singlePop_error}, we can establish a bound
specifically for the diagonals. If there exists a constant
$b_{3} \in (0,\infty )$ such that $L^{4} H \leq b_{3} \log (p)$ for
$N$ sufficiently large, which is natural since one would not expect
$H$ nor $L$ to grow with $N$, we then achieve essentially the same result
as in Theorem~\ref{thm:singlePop_error}: there exists a constant
$b_{4} \in (0, \infty )$ such that
$\|\widehat{\mbomega }- \mbomega ^{*}\|_{F} \leq b_{4} \sqrt{p \log (p)
/(\pi N)}$ with probability tending to one.

If each $\Omega ^{*}_{(h)}$ had a substantial number of zeros which were
not shared across all $H$ populations,~\eqref{eq:covEstimator} should perform
better with $\lambda > 0$. Specifically, we conjecture that in this case,
one could replace \textbf{A5} with a combination of \textbf{A2} (applied
to each population separately) and a relaxed version of \textbf{A5}, and
obtain an improved error bound relative to that in Theorem~\ref{thm:MultiPop_error} by taking $\lambda > 0$. However, proving this
type of result is technically challenging, and it is unclear whether some
of our intermediate results can be generalized.

\section{Numerical experiments}
\label{sec:Simulations}

\subsection{Data generating models and competing methods}
\label{sec6.1}

In this section, we compare the proposed estimator,~\eqref{eq:covEstimator}, to existing estimators under three data generating
models, Models 1--3, with $H = 4$ and
$(n,p) \in \{50, 100, 150\} \times \{40, 80, 120, 160, 200\}$ where
$n_{(1)} = \cdots = n_{(H)} = n$. In each replication, we generate
$\log (W_{(h)1}), \dots , \log (W_{(h)n_{(h)}})$ for $h \in [H]$ independently
with each $\log (W_{(h)i}) \in \mathbb{R}^{p}$ drawn from
${\mathrm{N}}_{p}(0, \Omega^* _{(h)})$.

The three models allow us to examine the methods' performance under different
types of shared sparsity. In Model 1, all covariance matrices are tridiagonal
with $\Omega^* _{(1)} = \Omega^* _{(2)}$ and
$\Omega^* _{(3)} = \Omega^* _{(4)}$. This is the ideal scenario for our method
since the sparsity patterns are identical across populations. In Model
2, only one $p/4 \times p/4$ diagonal block is nonzero in each covariance
matrix, though the block is in a different position for each
$\Omega^* _{(h)}$. Finally, in Model 3, $\Omega^* _{(1)}$ and
$\Omega^* _{(4)}$ do not share any nonzero off-diagonal elements, though
$\Omega^* _{(2)}$ and $\Omega^* _{(3)}$ share some nonzero off-diagonals
with each other, and with both $\Omega^* _{(1)}$ and
$\Omega^* _{(4)}$.

The specific models we consider are as follows.
\begin{itemize}
\item[] \textbf{Model 1.} The $\Omega^* _{(h)}$ are tridiagonal with either
all positive or all negative correlations:
\begin{equation*}
\Omega^* _{(h)jk} = \left \{
\begin{array}{rl}
0.3 \cdot \mathbf{1}(1 \leq |j - k| \leq 2) + \mathbf{1}(j = k) & :h
\in \{1, 2\}
\\
-0.2 \cdot \mathbf{1}(1 \leq |j - k| \leq 2) + \mathbf{1}(j = k) & :h
\in \{3, 4\}
\end{array}
\right .,
\end{equation*}
for $(h,j,k) \in [4] \times [p] \times [p]$.%
\smallskip
\item[] \textbf{Model 2.} The $\Omega^* _{(h)}$ are block diagonal with each
block having an AR(1) structure:
\begin{equation*}
\Omega^* _{(h)jk} = \left \{
\begin{array}{ll}
0.8^{|j-k|} & :|j-k| < p/4, ~(j,k) \in \mathcal{A}_{h}
\\
1 & :(j,k) \not \in \mathcal{A}_{h}, ~j = k
\end{array}
\right . ,~ (h,j,k) \in [4] \times [p] \times [p],
\end{equation*}
and $\mathcal{A}_{1} = [p/4] \times [p/4]$,
$\mathcal{A}_{2} = \{p/4 + 1, \dots , p/2\} \times \{p/4 + 1, \dots , p/2
\}$,
$\mathcal{A}_{3} = \{p/2 + 1, \dots , 3p/4\} \times \{p/2 + 1, \dots ,
3p/4\}$, and
$\mathcal{A}_{4} = \{3p/4 + 1, \dots , p\} \times \{3p/4 + 1, \dots , p
\}$.%
\smallskip
\item[] \textbf{Model 3.} The $\Omega^* _{(h)}$ have heterogeneous variances
and are block diagonal with diagonal blocks having an AR(1) structure,
i.e., $\Omega^* _{(h)} = D \Xi^*_{(h)} D$ where
\begin{equation*}
\Xi^*_{(h)jk} = \left \{
\begin{array}{ll}
0.9^{|j-k|} & :(j,k) \in \mathcal{B}_{h}
\\
1 & :(j,k) \not \in \mathcal{B}_{h}, ~j = k
\end{array}%
\right .,~ (h,j,k) \in [4] \times [p] \times [p],
\end{equation*}
$ D \in \mathbb{R}^{p \times p}$ is a diagonal matrix with diagonal entries
equally spaced from 3 to 1, and
$\mathcal{B}_{1} = [p/2] \times [p/2]$,
$\mathcal{B}_{2} = \{p/6 + 1, \dots , 2p/3\} \times \{p/6 + 1, \dots ,
2p/3\}$,
$\mathcal{B}_{3} = \{p/3 + 1, \dots , 5p/6\} \times \{p/3 + 1, \dots ,
5p/6\}$, and
$\mathcal{B}_{4} = \{p/2 + 1, \dots , p\} \times \{p/2 + 1, \dots , p
\}$
\end{itemize}
In order to select tuning parameters for each of the methods, we also generate
independent validation sets of the same size as the training set.

To estimate $\mbOmega ^{*}$, we consider multiple methods. All methods
but~\eqref{eq:covEstimator} estimate
$\Omega ^{*}_{(1)}, \dots , \Omega ^{*}_{(4)}$ separately. Specifically,
we use the method of \citet{cao2016large} with adapative soft-thresholding,
\texttt{COAT}, and use the method of \citet{fang2015cclasso},
\texttt{cclasso}. We also use an oracle estimator, \texttt{Oracle}, which
is the adaptively soft-thresholded sample covariance matrix of each
$\log (W_{(h)1}), \dots , \log (W_{(h)n_{(h)}})$. This is an oracle method
in the sense that we do not have access to the underlying abundances in
practice. Finally, we consider two versions of our method,
\texttt{SCC} and \texttt{SCC-H}, where \texttt{SCC} is short for
\underline{s}parse \underline{c}ompositional \underline{c}ovariance matrices.
The estimator $\texttt{SCC}$ is defined in~\eqref{eq:covEstimator}, whereas
$\texttt{SCC-H}$ is~\eqref{eq:SingleCovEstimator} with a separate tuning
parameter $\lambda $ chosen for each $h \in [H]$. The method
\texttt{SCC-H} estimates the covariances separately, but using a version
of our criterion. Including both \texttt{SCC} and \texttt{SCC-H} serves
to illustrate to what degree the improvement in performance is driven by
the loss function versus the sharing of sparsity patterns across the fibers
of $\mbOmega ^{*}$.

To assess the performance of each method, we measure the average (over
$H$ populations and 50 independent replications) Frobenius norm error and
$L_{1}$ matrix norm error of the estimated covariance matrices on the correlation
scale. We use the correlation scale because \texttt{cclasso} was designed
specifically for correlation matrix estimation. Relative performances under
both Frobenius norm and $L_{1}$ matrix norm error on the covariance scale
are similar and thus relegated to the Appendix.

We also measure true positive (TPR) and true negative rates (TNR) for each
method so that we may assess the recovery of nonzero correlations. Given
an estimate of $\mbOmega ^{*}$, $\widehat{\mbOmega}$, TPR and TNR are defined
as, respectively,
\begin{equation*}
\frac{1}{H}\sum _{h=1}^{H}
\frac{| \{ (j,k) : \bovtex{\widehat{\mbOmega}}_{hjk} \neq 0 \cap \bovtex{\mbOmega}^{*}_{hjk}\neq 0 \}|}{\left |\left \{ (j,k) : \bovtex{{\mbOmega}}^{*}_{hjk} \neq 0 \right \}\right |},~~~
\frac{1}{H}\sum _{h=1}^{H}
\frac{|\{ (j,k) : \bovtex{\widehat{\mbOmega}}_{hjk} = 0 \cap \bovtex{{\mbOmega}}^{*}_{hjk} = 0 \}|}{\left |\left \{(j,k) : \bovtex{{\mbOmega}}^{*}_{hjk} = 0 \right \}\right |}.
\end{equation*}

\subsection{Results}
\label{subsec:sim_results}

\begin{table}[t]
\centering
\scalebox{0.65}{
\begin{tabular}{rr|ccccc|ccccc|ccccc}
  \toprule
 && \multicolumn{5}{c|}{$n=50$} & \multicolumn{5}{c|}{$n=100$} & \multicolumn{5}{c}{$n=150$} \\
 && 40 & 80 & 120 & 160 & 200& 40 & 80 & 120 & 160 & 200  & 40 & 80 & 120 & 160 & 200  \\ 
  \midrule
\multicolumn{17}{c}{\textbf{Model 1}}\\
  \midrule
 \parbox[t]{2mm}{\multirow{5}{*}{\rotatebox[origin=c]{90}{TPR}}} &SCC & 0.953 & 0.928 & 0.904 & 0.890 & 0.875 & 0.999 & 0.998 & 0.996 & 0.996 & 0.996 & 1.000 & 1.000 & 1.000 & 1.000 & 0.999 \\ 
 & SCC-H  & 0.570 & 0.499 & 0.448 & 0.422 & 0.397 & 0.832 & 0.778 & 0.747 & 0.719 & 0.695 & 0.924 & 0.905 & 0.884 & 0.867 & 0.851 \\ 
 & COAT & 0.636 & 0.548 & 0.484 & 0.449 & 0.427 & 0.888 & 0.820 & 0.784 & 0.757 & 0.730 & 0.957 & 0.930 & 0.907 & 0.890 & 0.876 \\ 
 & Oracle & 0.653 & 0.567 & 0.498 & 0.464 & 0.438 & 0.888 & 0.821 & 0.788 & 0.760 & 0.733 & 0.951 & 0.927 & 0.905 & 0.889 & 0.875 \\ 
 & cclasso & 0.851 & 0.843 & 0.811 & 0.807 & 0.777 & 0.990 & 0.996 & 0.987 & 0.984 & 0.985 & 1.000 & 0.999 & 0.997 & 0.997 & 0.998 \\ 
  \midrule
 \parbox[t]{2mm}{\multirow{5}{*}{\rotatebox[origin=c]{90}{TNR}}} & SCC & 0.718 & 0.825 & 0.880 & 0.901 & 0.919 & 0.657 & 0.798 & 0.847 & 0.876 & 0.889 & 0.649 & 0.770 & 0.827 & 0.858 & 0.879 \\ 
 & SCC-H& 0.892 & 0.949 & 0.968 & 0.977 & 0.983 & 0.800 & 0.890 & 0.923 & 0.941 & 0.952 & 0.771 & 0.860 & 0.900 & 0.920 & 0.933 \\ 
 & COAT & 0.828 & 0.921 & 0.953 & 0.966 & 0.974 & 0.702 & 0.857 & 0.902 & 0.927 & 0.940 & 0.636 & 0.823 & 0.880 & 0.906 & 0.923 \\ 
 & Oracle & 0.858 & 0.924 & 0.953 & 0.966 & 0.973 & 0.777 & 0.874 & 0.909 & 0.929 & 0.942 & 0.762 & 0.850 & 0.890 & 0.912 & 0.926 \\ 
 & cclasso & 0.231 & 0.281 & 0.374 & 0.395 & 0.465 & 0.023 & 0.028 & 0.085 & 0.104 & 0.111 & 0.001 & 0.007 & 0.024 & 0.035 & 0.054 \\ 
   \midrule
\multicolumn{17}{c}{\textbf{Model 2}}\\
  \midrule
 \parbox[t]{2mm}{\multirow{5}{*}{\rotatebox[origin=c]{90}{TPR}}} &   SCC  & 0.904 & 0.709 & 0.569 & 0.489 & 0.414 & 0.954 & 0.798 & 0.653 & 0.554 & 0.489 & 0.970 & 0.821 & 0.697 & 0.589 & 0.522 \\ 
&  SCC-H & 0.766 & 0.529 & 0.404 & 0.333 & 0.277 & 0.856 & 0.631 & 0.486 & 0.401 & 0.340 & 0.883 & 0.672 & 0.529 & 0.434 & 0.374 \\ 
 & COAT & 0.805 & 0.607 & 0.468 & 0.388 & 0.320 & 0.886 & 0.747 & 0.590 & 0.481 & 0.400 & 0.921 & 0.810 & 0.678 & 0.548 & 0.461 \\ 
 & Oracle & 0.897 & 0.632 & 0.478 & 0.391 & 0.326 & 0.953 & 0.707 & 0.544 & 0.443 & 0.379 & 0.975 & 0.743 & 0.580 & 0.476 & 0.407 \\ 
 & cclasso &  & 0.999 & 0.997 & 0.995 & 0.989 & 1.000 & 1.000 & 1.000 & 0.999 & 0.999 & 1.000 & 1.000 & 1.000 & 1.000 & 1.000 \\
  \midrule
  \parbox[t]{2mm}{\multirow{5}{*}{\rotatebox[origin=c]{90}{TNR}}} & SCC & 0.238 & 0.481 & 0.605 & 0.663 & 0.722 & 0.136 & 0.374 & 0.532 & 0.618 & 0.661 & 0.098 & 0.350 & 0.490 & 0.583 & 0.630 \\ 
&  SCC-H & 0.536 & 0.718 & 0.792 & 0.827 & 0.858 & 0.462 & 0.644 & 0.749 & 0.797 & 0.824 & 0.436 & 0.625 & 0.725 & 0.776 & 0.807 \\ 
&  COAT & 0.326 & 0.586 & 0.715 & 0.774 & 0.819 & 0.196 & 0.412 & 0.601 & 0.706 & 0.764 & 0.137 & 0.329 & 0.511 & 0.638 & 0.714 \\ 
 & Oracle& 0.576 & 0.704 & 0.765 & 0.799 & 0.831 & 0.544 & 0.667 & 0.737 & 0.782 & 0.808 & 0.521 & 0.645 & 0.726 & 0.768 & 0.795 \\ 
&  cclasso&  & 0.002 & 0.004 & 0.006 & 0.013 & 0.000 & 0.000 & 0.000 & 0.001 & 0.002 & 0.000 & 0.000 & 0.000 & 0.000 & 0.000 \\ 
  \midrule
\multicolumn{17}{c}{\textbf{Model 3}}\\
  \midrule
 \parbox[t]{2mm}{\multirow{5}{*}{\rotatebox[origin=c]{90}{TPR}}} & SCC  & 0.933 & 0.732 & 0.623 & 0.550 & 0.488 & 0.984 & 0.804 & 0.696 & 0.616 & 0.545 & 0.992 & 0.880 & 0.761 & 0.655 & 0.597 \\ 
 & SCC-H& 0.696 & 0.556 & 0.463 & 0.404 & 0.342 & 0.794 & 0.654 & 0.567 & 0.492 & 0.430 & 0.831 & 0.696 & 0.618 & 0.534 & 0.469 \\ 
 & COAT & 0.838 & 0.740 & 0.638 & 0.556 & 0.470 & 0.920 & 0.869 & 0.803 & 0.715 & 0.628 & 0.942 & 0.920 & 0.873 & 0.810 & 0.731 \\ 
 & Oracle  & 0.960 & 0.772 & 0.620 & 0.524 & 0.445 & 0.988 & 0.830 & 0.700 & 0.596 & 0.504 & 0.994 & 0.872 & 0.726 & 0.619 & 0.544 \\ 
 & cclasso &  & 0.999 & 0.998 & 0.997 & 0.994 &  & 1.000 & 1.000 & 0.999 & 0.999 & 1.000 & 1.000 & 1.000 & 1.000 & 1.000 \\ 
  \midrule
  \parbox[t]{2mm}{\multirow{5}{*}{\rotatebox[origin=c]{90}{TNR}}} & SCC & 0.106 & 0.395 & 0.521 & 0.598 & 0.640 & 0.025 & 0.324 & 0.467 & 0.555 & 0.616 & 0.011 & 0.210 & 0.396 & 0.519 & 0.571 \\ 
 & SCC-H & 0.433 & 0.652 & 0.744 & 0.791 & 0.825 & 0.322 & 0.561 & 0.668 & 0.738 & 0.784 & 0.292 & 0.518 & 0.639 & 0.713 & 0.759 \\  
 & COAT & 0.288 & 0.514 & 0.630 & 0.701 & 0.750 & 0.137 & 0.333 & 0.462 & 0.575 & 0.653 & 0.089 & 0.225 & 0.369 & 0.477 & 0.565 \\ 
 & Oracle & 0.560 & 0.647 & 0.697 & 0.745 & 0.773 & 0.515 & 0.617 & 0.662 & 0.708 & 0.757 & 0.543 & 0.587 & 0.657 & 0.707 & 0.745 \\ 
 & cclasso &  & 0.001 & 0.002 & 0.003 & 0.006 &  & 0.000 & 0.000 & 0.000 & 0.001 & 0.000 & 0.000 & 0.000 & 0.000 & 0.000 \\ 
\bottomrule
\end{tabular}
}
\caption{True positive and true negative rates for each of the methods averaged over 50 independent replications under Model 1--3.}
\end{table}

\begin{figure}[t!]
\begin{center}
\includegraphics[width=0.95\textwidth]{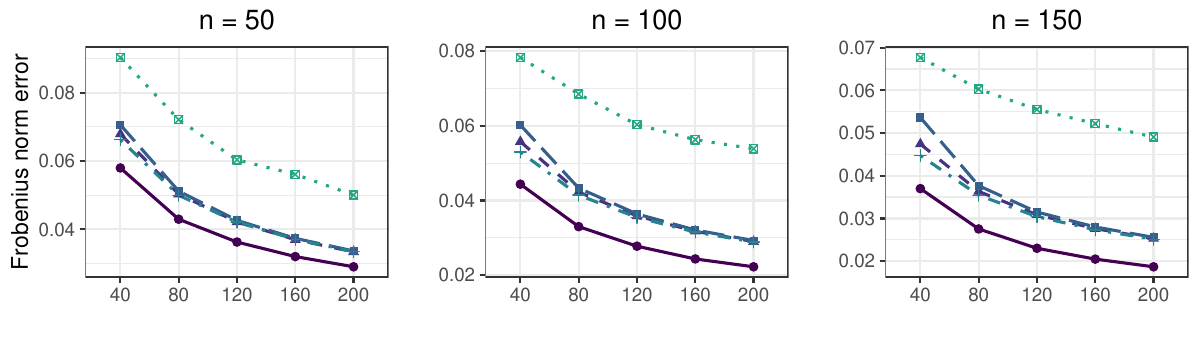}\\
\includegraphics[width=0.95\textwidth]{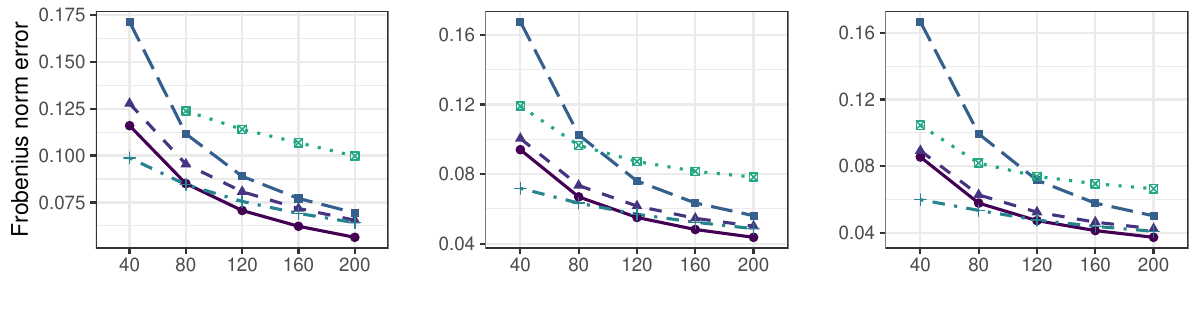}\\
\includegraphics[width=0.95\textwidth]{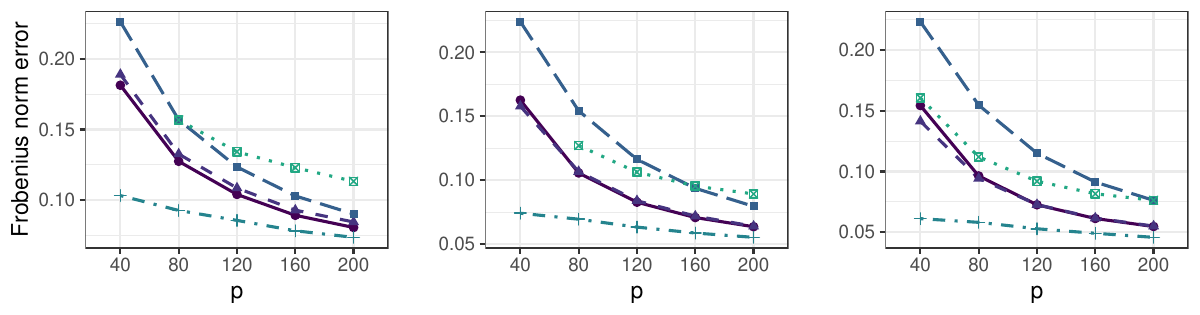}\\
\includegraphics[width=10cm]{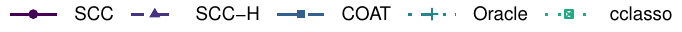}
\end{center}
\vspace*{-12pt}
\caption{Average Frobenius norm error divided by $p$ (on the correlation
scale) over 50 independent replications under (top row) Model 1, (middle
row) Model 2, and (bottom row) Model 3 with
$(n,p) \in \{50, 100, 150\} \times \{40, 80, 120, 160, 200\}$.}
\label{fig:FrobCor}
\end{figure}

In Figure~\ref{fig:FrobCor}, we display average Frobenius norm errors (divided
by $p$). With $p = 40$, the \texttt{cclasso} software would sometimes return
undefined estimates (\texttt{NA} in R), so we omit comparisons in these
settings. Unsurprisingly, under Model 1, \texttt{SCC} substantially outperforms
all of the competitors, including \texttt{Oracle}. This illustrates the
utility of exploiting shared sparsity patterns when estimating multiple
covariance matrices. Notably, \texttt{Oracle}, \texttt{COAT}, and
\texttt{SCC-H} all perform similarly in each setting under Model 1. Under
Model 2, \texttt{SCC} outperforms all competitors, including
\texttt{Oracle}, once $p \geq 120$. Comparing the competitors which could
be used in practice, \texttt{SCC-H} performs better than
\texttt{COAT} and \texttt{cclasso} in all situations. The estimator
\texttt{COAT} performs worse than \texttt{cclasso} for small $p$, but significantly
outperforms \texttt{cclasso} once $p \geq 120$. Finally, under Model 3,
\texttt{SCC} and \texttt{SCC-H} perform similarly in every setting. The
only method to outperform \texttt{SCC} and \texttt{SCC-H} is
\texttt{Oracle}, which cannot be used in practice. The fact that
\texttt{Oracle} outperforms the other methods so substantially speaks to
the difficulty of estimating the covariances under this data generating
model relative to Model 1 and 2.

\begin{figure}[t!]
\begin{center}
\includegraphics[width=0.95\textwidth]{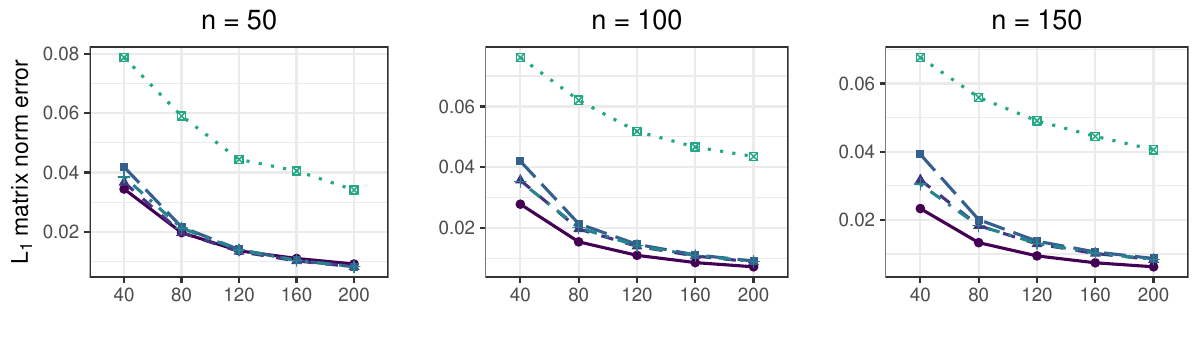}\\
\includegraphics[width=0.95\textwidth]{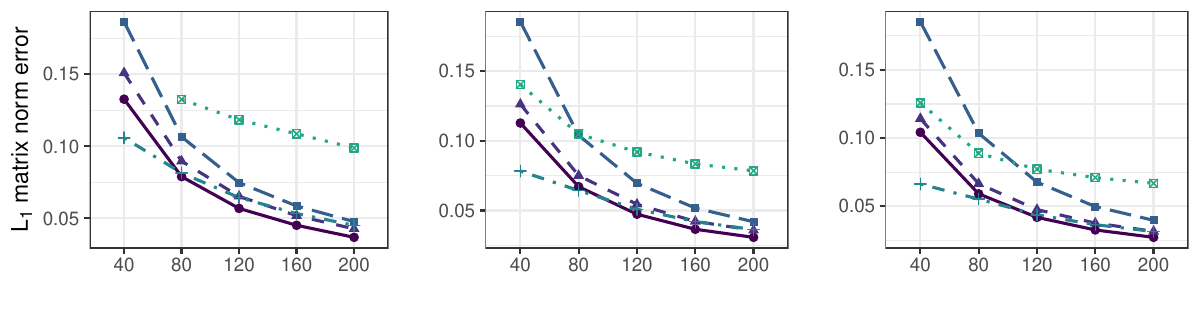}\\
\includegraphics[width=0.95\textwidth]{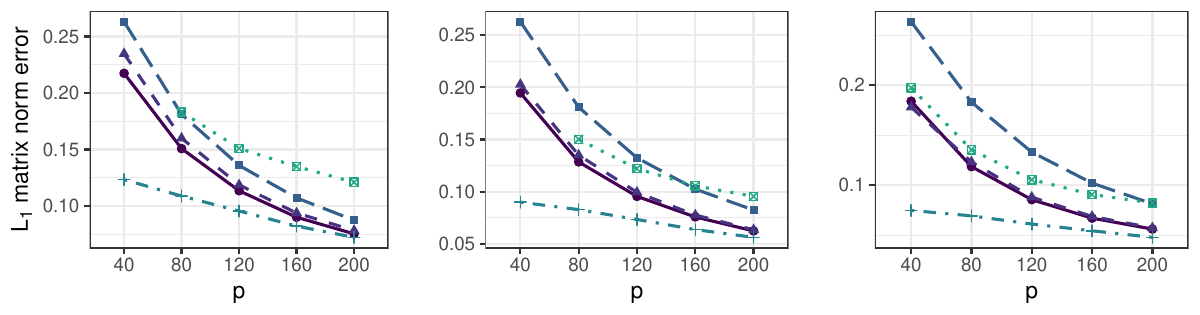}
\includegraphics[width=10cm]{r1_plots/Legend.pdf}
\end{center}
\vspace*{-12pt}
\caption{Average $L_{1}$ matrix norm error divided by $p$ (on the correlation
scale) over 50 independent replications under (top row) Model 1, (middle
row) Model 2, and (bottom row) Model 3 with
$(n,p) \in \{50, 100, 150\} \times \{40, 80, 120, 160, 200\}$.}
\label{fig:L1Cor}
\end{figure}

One should be careful drawing conclusions based on Frobenius norm error
results alone, however, because our methods, \texttt{SCC} and
\texttt{SCC-H}, both minimize a Frobenius norm criterion, whereas
\texttt{COAT} does not. Thus, these results may be biased in favor of
\texttt{SCC} and \texttt{SCC-H}. For this reason, we also included
$L_{1}$ matrix norm results in Figure~\ref{fig:L1Cor}. Here, the
$L_{1}$ matrix norm is the maximum of the $L_{1}$ vector-norm of the columns
of a matrix. Under Model 1 with $n=50$, there appears to be little difference
between the methods---other than \texttt{cclasso}---in terms of
$L_{1}$ matrix norm. However, when $n \geq 100$, \texttt{SCC} significantly
outperforms all competitors. Under Model 2, the results more closely mirror
those in Figure~\ref{fig:FrobCor}: \texttt{SCC} outperforms all competitors,
including \texttt{Oracle}, when $p \geq 120$. The results under Model 3,
relatively speaking, are similar to those observed under Model 3 using
Frobenius norm error. The method \texttt{Oracle} performs best, but among
the methods which could be used in practice, \texttt{SCC} and
\texttt{SCC-H} clearly outperform \texttt{cclasso} and \texttt{COAT}.

The performance of \texttt{SCC} and \texttt{SCC-H} can be partially explained
by their performance in recovering the true set of nonzero off-diagonals.
In Model 1, \texttt{SCC} has nearly perfect TPR, and TNR only slightly
lower than the best performing competitor. \texttt{SCC-H} tends to have
similar TPR as \texttt{COAT}, but also tends to have higher TNR. A similar
conclusion can be drawn under Model 2. Under Model 3, however,
\texttt{COAT} tends to have higher TPR and similar TNR to
\texttt{SCC}, whereas \texttt{SCC-H} has lower TPR and higher TNR than
\texttt{COAT}. Note that under Model 3, \texttt{oracle} does well in part
due to the fact that the covariances have varying diagonals.

In the Appendix, we provide additional
simulation study results. First, we present the results from Figures~\ref{fig:FrobCor} and~\ref{fig:L1Cor}, but on the covariance scale. Relative
performances closely mirror those in Figures~\ref{fig:FrobCor} and~\ref{fig:L1Cor}. Second, we present results comparing \texttt{SCC-H} to
\texttt{COAT} and \texttt{cclasso} in terms of estimating
$\Omega _{(1)}^{*}$ under Models 1--3. Our results clearly demonstrate
that \texttt{SCC-H} can significantly outperform both \texttt{COAT} and
\texttt{cclasso} for single population basis covariance matrix estimation.
Finally, we also perform additional simulation studies wherein the sample
sizes $(n_{(1)}, n_{(2)}, n_{(3)}, n_{(4)}) = (100, 75, 50, 25)$. We compare
the same competitors as in Section~\ref{subsec:sim_results}, but also include~\eqref{eq:covEstimator_weighted}. Under Model 1,~\eqref{eq:covEstimator_weighted} outperforms the competitors, though under
Models 2 and 3, there is little difference between~\eqref{eq:covEstimator} and~\eqref{eq:covEstimator_weighted}.

\section{Analysis of microbiome in myalgic encephalomyelitis/chronic fatigue syndrome}
\label{sec:real_data_anaylsis}

\subsection{Basis covariance matrix estimation}
\label{sec7.1}

We illustrate our method by analyzing data on the gut microbiome of patients
diagnosed with myalgic encephalomyelitis/chronic fatigue syndrome (ME/CFS) versus
controls from \citet{Giloteaux2016}. In order to obtain the microbial profiles,
\citet{Giloteaux2016} sequenced 16S rRNA genes from stool samples using
Illumina MiSeq. After first filtering patients based on total reads ($
\geq 5000$), we filter operational taxonomic units (OTUs) to only those
that comprise at least 10\% of total reads in one or more patients. This
reduced the original 138 OTUs to $p = 39$ OTUs, though led to us filtering
out only 8\% and 10\% of each subjects' total reads (on average) in controls
and ME/CFS, respectively. Following \citet{cao2016large}, we add 0.5 to
all counts to avoid zeros before converting counts to compositions. To
be clear, these counts $Y_{(h)i}$ are distinct from the latent abundances
$W_{(h)i}$, which are assumed to be independent and identically distributed
for all $k \in [n_{(h)}]$. For example, we may assume that
$Y_{(h)ij} = M_{(h)i} W_{(h)ij}$ where $M_{(h)i}$ is a positive random
variable \citep{ma2021networks}, so that
$Y_{(h)ij}/\sum _{k=1}^{p} Y_{(h)ik} = X_{(h)ij} = W_{(h)ij}/\sum _{k=1}^{p}
W_{(h)ik}$. In the Appendix, we demonstrate
that our estimates do not change much when we use a pseudocount of 0.01
instead of 0.5 to handle zeros.

\begin{figure}
\begin{center}
\includegraphics[width=6.0cm]{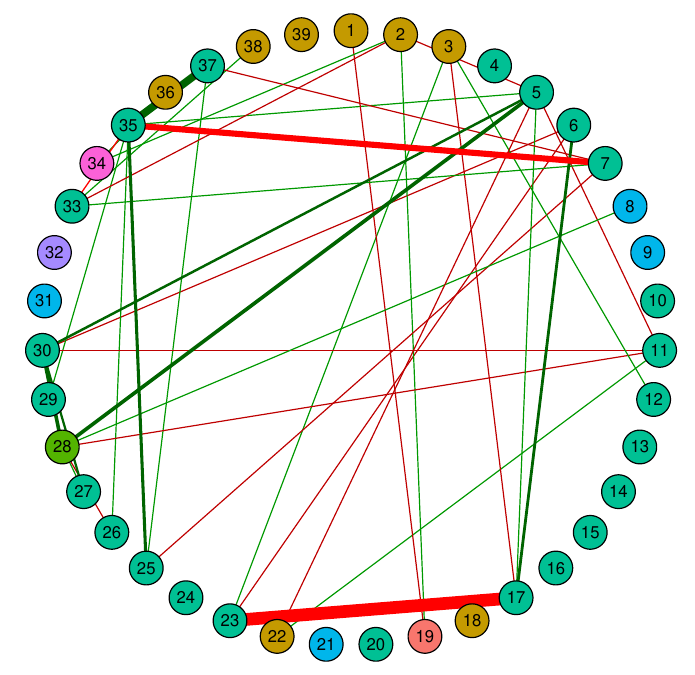}~~~~~~
\includegraphics[width=7.6cm]{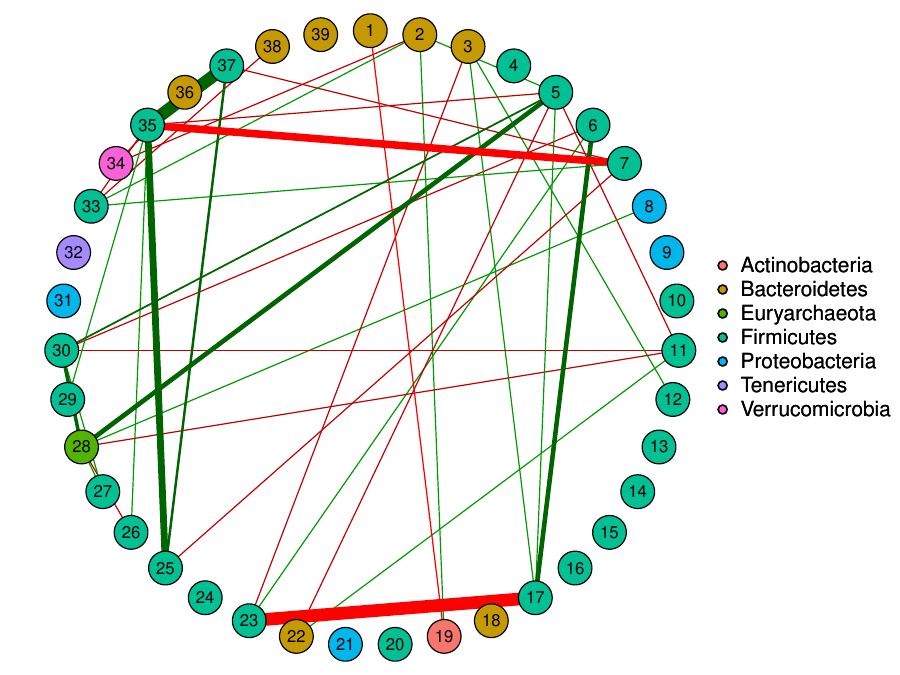}\\
(a) Controls ~~~~~~~~~~~~~~~~~~~~~~~~~~~~~ ~~~(b) ME/CFS~~~~~~~~~~~~~~\\
\caption{Estimated correlation networks using \eqref{eq:covEstimator} for (a) control patients and (b) ME/CFS patients. The thickness of the edge corresponds to the strength of the association: stronger associations are represented by thicker edges. Positive and negative correlations are colored, respectively, with green and red, while a zero correlation is represented by the lack of an edge. }\label{fig:estimated_covariances_chronicfatigue}
\end{center}
\end{figure}

Our estimates of the covariance matrices, with tuning parameters chosen
using ten-fold cross-validation, are in Figure~\ref{fig:estimated_covariances_chronicfatigue}. Each node in these graphs
represents a unique OTU. The nodes are colored according to OTU's phylum
and each node's family, genus, and species is provided in the Appendix. The thickness of the edge corresponds to the strength of
the association: stronger associations are represented by thicker edges.
Positive and negative correlations are colored, respectively, with green
and red, while a zero correlation is represented by the absence of an edge.

Examining the estimated covariance matrices, the majority of associations
occur within two OTUs belonging to the same phylum. We also see that our
method estimates the two groups' covariance matrices to have identical
sparsity patterns, in sharp contrast with the estimates based on
\texttt{COAT}, the method of \citet{cao2016large} (Figure~\ref{fig:estimated_covariances_chronicfatigue_coat}). Most strong positive
and negative associations are shared across the two groups. Notably, one
of the eight associations whose direction differ across controls and ME/CFS
is (23--3; \textit{Ruminococcus bromii}--\textit{Bacteroides ovatus}).
\textit{Ruminococcus bromii} is known to degrade resistant starch particles
inaccessible to other bacteria, whereas \textit{Bacteroides ovatus} digests
inulin \citep{porter2016love}. That these two OTUs are estimated to be
associated is interesting since both resistant starch and inulin are fermentable
carbohydrates whose joint behavior has been of interest in past studies
\citep{younes2001effects}.

In addition, an insight gleaned from our estimates is that more negative
associations are observed in chronic fatigue syndrome patients than in
controls. This coheres with the reduced diversity in the microbiome communities
for ME/CFS patients observed by \citet{Giloteaux2016}. Moreover, the positive
associations between (25--35) and (6--17) are much stronger in ME/CFS than
in controls. This too suggests reduced diversity in ME/CFS as OTUs labeled 6, 35, 25,
and 17 all belong to the same phylum, \textit{Firmicutes}.

Estimates using \texttt{COAT} are more difficult to interpret. First, there
is a larger number of nonzero entries in both estimates, and their sparsity
patterns differ substantially. In total, \texttt{COAT} identifies 185 associations
in one population not present in the other. Moreover, the estimates from
\texttt{COAT} disagree in terms of their strongest associations. For example,
one of the strongest positive associations estimated in controls is between
(29--13), whereas in patients with ME/CFS, their method estimates these
two OTUs to be uncorrelated.

Finally, we emphasize that our estimator~\eqref{eq:covEstimator} does not
require that sparsity patterns are identical across CFS and controls. Instead,
the similarity of sparsity patterns is determined by the combination of
tuning parameters $(\gamma , \lambda )$, which are selected by cross-validation.
Thus, in this application, it is the data which suggest that the sparsity
patterns are identical.

\subsection{Stability assessment}
\label{subsec:stability}

We perform a stability assessment to determine to what degree our respective
estimates, displayed in Figures~\ref{fig:estimated_covariances_chronicfatigue_coat} and~\ref{fig:estimated_covariances_chronicfatigue}, are reliable. Following
\citet{cao2016large}, we generate 100 independent bootstrap samples and
refit both estimators to the bootstrapped samples. We say an estimated
nonzero correlation is stable if it is nonzero in at least 80 of the 100
bootstrap samples. In Table~\ref{tab:stability}, we report the stability
of each correlation estimate.

In the first four columns, we assess the stability of all correlations:
in rows labeled positive and negative, we report the number of correlations
estimated to be positive and negative, respectively, in the estimates displayed
in Figures~\ref{fig:estimated_covariances_chronicfatigue_coat} and~\ref{fig:estimated_covariances_chronicfatigue}. In the row labeled stability,
we report the percentage of these correlations which were estimated to
be nonzero at least 80 of the 100 bootstrap samples. For example,
\texttt{SCC} estimated 22 positive and 16 negative correlations (Figure~\ref{fig:estimated_covariances_chronicfatigue}a); of these 38 correlations,
89.5\% of them were estimated to be nonzero in at least 80 bootstrap samples.
For our method, in both controls and ME/CFS basis covariance matrix estimates,
almost all of the edges we estimated to be nonzero are stable.
\texttt{COAT}, on the other hand, has lower stability in both controls
and ME/CFS.

In the first row of the ``shared correlations'' columns of Table~\ref{tab:stability}, we report the number of estimated correlations where
a correlation was positive in both estimates (controls and ME/CFS), or
negative in both estimates. Our method estimated that all correlations
are shared, and has reasonably high stability. In particular, this column
suggests that of the 40 shared correlations, 90\% were estimated in at
least 80 bootstrap samples. \texttt{COAT} has slightly lower stability
for its shared correlations despite estimating fewer than half as many
as our method.

\begin{table}[t]
\tabcolsep=5pt

\caption{Stability for all correlations, shared correlations, and distinct
correlations over 100 bootstrap samples. For the distinct correlation columns,
D1 refers to a correlation which was nonzero in controls, but zero in ME/CFS,
whereas D2 refers to a correlation which was zero in controls but nonzero
in ME/CFS.}
\label{tab:stability}
\centering
\scalebox{0.82}{
\begin{tabular}{rcc|cc|ccc|ccc}
  \toprule
  &  \multicolumn{4}{c|}{All correlations} & \multicolumn{3}{c|}{Shared correlations} & \multicolumn{3}{c}{Distinct correlations}\\
   \midrule
  & \multicolumn{2}{c}{\texttt{SCC}} & \multicolumn{2}{c|}{\texttt{COAT}}&&& &&&\\
 & Control\ & \hspace{-3pt} ME/CFS \hspace{-2pt}& Control\ & \hspace{-3pt}ME/CFS\hspace{-2pt} & & \texttt{SCC} & \texttt{COAT} & &\texttt{SCC} & \texttt{COAT}\\
  \midrule
Positive & 22 & 21 & 53 & 30 & Same sign & 29 & 19 & D1 &  0 & 111 \\
Negative & 16 & 17 & 82 & 68 & Diff. sign & 9 & 5 & D2 & 0 & 74\\
 Stability & 89.5\% & 86.8\% & 83.0\%& 84.7\%  & Stability & 86.8\%  & 58.3\% & Stability & --- & 00.0\% \\
   \bottomrule
\end{tabular}
}
%
\end{table}

Finally, the most telling result comes in the ``distinct correlations''
columns of Table~\ref{tab:stability}. Here, we report the number of correlations
which were nonzero in controls and zero in ME/CFS (D1) and the number of
correlations which were zero in controls and nonzero in ME/CFS (D2). We
see that \texttt{SCC} estimates no correlations to be distinct, whereas
\texttt{COAT} estimates 185 correlations to be distinct. However, the stability
of these correlations is zero: none of these distinct correlations appeared
in 80 or more of the bootstrap samples. These results suggest that the
estimates provided by our method may be more reliable than
\texttt{COAT}.

\section{Discussion}
\label{sec8}

In this article, we proposed a new method for estimating basis covariance
matrices from compositional data. An important question about our method
is whether it could provide reasonable estimates of the basis precision
(inverse covariance) matrix. Though our method can provide
estimates of $\Omega _{*(h)}^{-1}$ (since our estimates are always positive
definite), these estimates will not, in general, be sparse. If a practitioner is
interested in sparse precision matrix estimation, we recommend using methods
specifically designed for this task, e.g., \citet{zhang2023care}. To the
best of knowledge, there exist no methods for jointly estimating multiple
sparse precision matrices from compositional data. This could be a fruitful
direction for future research.

There are two aspects of our data analysis which could be improved. First,
the original data were counts (reads per OTU), which we converted to compositions.
It has been argued that total reads per patient is an experimental artifact,
and thus, microbiome sequencing data should be converted to compositions
\cite[e.g., see][and references therein]{gloor2017microbiome}. However,
as pointed out by a referee, 
there is nonetheless some loss of information when we ignore total reads
per patient. Ideally, an estimator could somehow make use of this additional
information. Second, our method assumes that components of the observed
composition are positive with probability one. In 16S rRNA sequencing (microbiome)
data, however, it is common to observe many zeros. Thus, as future work,
we hope to extend our method to address these two issues.

\subsection*{Acknowledgements}
Aaron J. Molstad was supported in part by NSF DMS-2113589. Piotr M. Suder was supported in part by University Scholars Program
at the University of Florida.


\bibliography{CompositionalCovariance}

\end{document}